\documentclass[%
reprint,
 amsmath,amssymb,
 aps,
floatfix,
]{revtex4-2}

\usepackage{graphicx}
\usepackage{dcolumn}
\usepackage{bm}
\usepackage{hyperref}
\usepackage{booktabs}

\usepackage{quantikz}
\usepackage{tikz}
\usepackage{float}
\usepackage{enumitem}
\usepackage{makecell}

\begin{document}

\preprint{APS/123-QED}

\title{Pulsar Classification: Comparing Quantum Convolutional Neural Networks and Quantum Support Vector Machines}

\author{Donovan Slabbert}
 \email{donovanslab@mweb.co.za}
\author{Matt Lourens}
 \email{lourensmattj@gmail.com}
\affiliation{Department of Physics, Stellenbosch University, Stellenbosch, South Africa \\
 }

\author{Francesco Petruccione}
 \email{petruccione@sun.ac.za}
\affiliation{National Institute of Theoretical and Computational Sciences (NITheCS), Stellenbosch, South Africa \\
 School of Data Science and Computational Thinking, Stellenbosch University, Stellenbosch, South Africa \\
 Department of Physics, Stellenbosch University, Stellenbosch, South Africa \\
}

\date{\today}

\begin{abstract}
Well-known quantum machine learning techniques, namely quantum kernel assisted support vector machines (QSVMs) and quantum convolutional neural networks (QCNNs), are applied to the binary classification of pulsars. In this comparitive study it is illustrated with simulations that both quantum methods successfully achieve effective classification of the HTRU-2 data set that connects pulsar class labels to eight separate features. QCNNs outperform the QSVMs with respect to time taken to train and predict, however, if the current NISQ era devices are considered and noise included in the comparison, then QSVMs are preferred. QSVMs also perform better overall compared to QCNNs when performance metrics are used to evaluate both methods. Classical methods are also implemented to serve as benchmark for comparison with the quantum approaches.
\end{abstract}

                              
\maketitle


\section{\label{sec:Introduction} Introduction \protect}

In the field of astronomy, where vast amounts of observational data of various types of astronomical objects are collected annually, cataloging these objects into their respective categories is crucial. This comparative study focuses on pulsars, which are a type of neutron star formed when a massive star reaches the end of its life, collapsing under gravity but not enough to overcome neutron degeneracy pressure \cite{lyne2012pulsar, kippenhahn1990stellar, maoz2016astrophysics}. This results in a dense star remnant with a significantly smaller radius and rapid rotation, often on the order of seconds to milliseconds \cite{manchester1977pulsars}. Pulsars possess strong magnetic fields that emit particle beams from their poles. If Earth happens to intercept these beams we observe periodic pulses of observable signals, typically measured using radio telescopes. These neutron stars exhibiting this periodic signal are known as pulsars.

Identifying pulsars is essential for distinguishing them from typical main sequence stars and other neutron stars, as binary systems consisting of two pulsars can generate gravitational waves \cite{hulse1975discovery, stairs2003testing}. Pulsars' precise rotational periods also make them useful for detecting gravitational waves passing between them and Earth, by observing deviations in their pulses \cite{foster1990constructing, mclaughlin2013north, verbiest2016international}. This capability enhances our ability to study gravitational waves.

With the increasing volume of astronomical data due to improving observational technology, traditional software will eventually struggle to manage and process it effectively. The Square Kilometer Array (SKA), set to launch soon, is expected to produce data on the order of exabytes ($10^{18}$ bytes) \cite{an2019science}. Real-time data analysis and efficient classification tasks are the ultimate goals in astronomy and data science. While current classical classification methods suffice for now, the problem will become insurmountable as data sizes continue to grow.

It is for this reason that machine learning has been thoroughly explored as a solution to this challenge. Classical machine learning has been applied to pulsar classification \cite{wang2019hybrid, tariq2022adaboost, beniwal2021detection, xiao2020pulsar, lee2012application, wang2019pulsar, lyon2016fifty}, however, since classical approaches face limitations with increasing data sizes \cite{zhou2017machine}, the problem might have to be approached with a different perspective. Quantum computing (QC), particularly quantum machine learning (QML) that combines the computational power of quantum computers with machine learning, is such a new perspective. We endeavour to explore the applicability of quantum computers to the machine learning task of classifying pulsars. We do this by comparing two quantum approaches. Quantum methods (one of which is also called a QSVM) have been used on the HTRU-2 data set before \cite{sarma2019quantum}. Another known quantum approach \cite{kordzanganeh2021quantum} uses a circuit with a depth that would introduce a lot of noise.

We improve over the accuracy presented in this case \cite{kordzanganeh2021quantum} with shorter depth circuits. Our study uses QML approaches that have circuits with lower depths to effectively and accurately classify pulsars using the real-world HTRU-2 data set using noise free and noisy simulated quantum algorithms and in a limited implementation on currently available NISQ devices. The HTRU-2 data has been manually created by experts from a large number of pulsar candidates \cite{misc_htru2_372}. This means that the problem falls under the machine learning category of supervised machine learning \cite{mohri2018foundations, alpaydin2020introduction} in that we have a ground truth to use in training, but this ground truth was decided on by human experts. There are eight features present in the data set for each candidate. Each of the eight features are encoded into a quantum state using rotational encoding. The quantum approaches under consideration are a basic 8-qubit implementation of a quantum-assisted support vector machine (QSVM) and an 8-qubit quantum convolutional neural network (QCNN). Both approaches will be compared with each other as well as to their classical counterparts.

The two methods implemented are variants of commonly used QML archetypes, namely quantum kernel methods and variational quantum algorithms. The goal is to investigate which is currently more suited for real-world binary classification and the variants of choice to investigate are the QSVM and QCNN. In \cite{schuld2021supervised} the assertion is made that variational quantum algorithms can also be considered to be quantum kernel methods, but the difference in this variational approach is the use of hierarchical ideas inspired by classical convolutional neural networks and the training procedure. The QSVM evaluates the kernel using a quantum device, followed by fitting a classical SVM, whereas the QCNN relies on a classical optimization loop to optimize parameters used in its parameterized circuit. There are theoretical guarantees associated with kernel methods \cite{schuld2021supervised} and it is these guarantees that we want to compare the QCNN with using its more manipulable quantum circuit. 

We show below how the QCNN, despite its larger circuit, outperforms the QSVM when considering the time taken to train the model and to use it for prediction. Using standard confusion matrices \cite{muller2016introduction} and evaluation metrics \cite{powers2020evaluation} the best method is not as apparent. The QSVM generally performs slightly better depending on the metric used, however, of the two methods the QSVM avoids the most false negatives. In this discovery-focused use case, avoiding false negatives is really important. A short noise test illustrates the obvious point that longer quantum circuits undergo more noise. This is because there is an error associated with each quantum logic gate applied \cite{nielsen2001quantum}. With noise included in the comparison, the QSVM performs better because of its shorter depth.

The rest of the paper is structured as follows: Section 2 is an explanation of binary classification as a supervised machine learning task. It also contains the necessary information regarding the two methods that are implemented. Section 3 is an explanation of how the HTRU-2 data set is used and how data preprocessing was performed. Here the specific methodology used for both approaches are also discussed Section 4 serves as a comprehensive comparison between the two methods. This includes noise free and noisy simulated results as well as limited real quantum device results. Section 5 is a short conclusion discussion.

\section{\label{sec:Theory}Theory \protect}

\subsection{\label{sec:Binary}Binary Classification}

Binary classification is a supervised machine learning problem where samples in a data set are categorized into one of two classes, in this case as being a pulsar or not. The following is the formal problem statement for binary classification \cite{schuld2021machine}:

\begin{equation}
    f_{\theta}:\bm{X} \rightarrow \bm{Y}, f_{\theta}(\vec{x}) = y, \vec{x} \in \bm{X} \, \text{and} \, y \in \bm{Y},
\end{equation}

\noindent where $\bm{X}$ and $\bm{Y}$ are the input and output domain respectively, $x$ is a $n\text{-dimensional}$ input vector (also called a feature vector) that contains $n$ entries or features, and $y$ is a class label from the set $\{0, 1\}$, where a label of $1$ indicates the positive class (pulsars) and $0$ the negative class. The function $f_{\hat{\theta}}$ is the optimal model from a model family, or set of models $\{f_{\theta}\}$, and the subscript $\theta$ indicates that this function has a set of parameters that need to be optimised during training.

Training involves the use of a subset of both $\bm{X}$ and $\bm{Y}$, called the training set $\bm{X_{\text{train}}}$ and $\bm{Y_{\text{train}}}$, during an optimisation or fitting process to find the best model from a model family that predicts the testing set $\bm{X_{\text{test}}}$ and $\bm{Y_{\text{test}}}$ labels as accurately as possible. Based on the predictions made, cross-referenced with the original labels, the output can be formulated as a confusion matrix, which is an $2 \times 2$ array containing the prediction outcomes. Confusion matrices are useful for assessing how well a model predicts on new samples by visually representing the distribution of predictions. Once a model has been trained that performs at a level that is required, the model can be used to predict on previously unseen data, where no prior knowledge of the labels are possible.

\begin{table}[t]
\centering
\renewcommand{\arraystretch}{1.4}
\setlength{\tabcolsep}{4pt}
\begin{tabular}{ll} 
\toprule
\textbf{Metric} & \textbf{Equation} \\
\midrule
\text{Accuracy} & $\frac{\text{TP} + \text{TN}}{\text{TP} + \text{TN} + \text{FP} + \text{FN}}$ \\
\text{Recall} & $\frac{\text{TP}}{\text{TP} + \text{FN}}$ \\
\text{Specificity} & $\frac{\text{TN}}{\text{TN} + \text{FP}}$ \\
\text{Precision} & $\frac{\text{TP}}{\text{TP} + \text{FP}}$ \\
\text{Negative Prediction Value} & $\frac{\text{TN}}{\text{TN} + \text{FN}}$ \\
\text{Balanced Accuracy} & $\frac{1}{2}(\text{Recall} + \text{Specificity})$ \\
\text{Geometric Mean} & $\sqrt{\text{Recall} \cdot \text{Specificity}}$ \\
\text{Informedness} & $\text{Recall} + \text{Specificity} - 1$ \\
\bottomrule
\end{tabular}
\caption{The expressions for the evaluation metrics used. TP, TN, FP, and FN represent the true positives, true negatives, false positives, and false negatives respectively after using a binary classification model on a testing set. These values form part of confusion matrices.}
\label{table=equations}
\end{table}

Various metrics are chosen to quantify how well a model performs at prediction and are calculated using values extracted from the confusion matrices. In our case it is more important to classify possible pulsar candidates instead of finding true non-pulsars. This is to aid the discovery of new pulsars, which implies that true positives, false positives and false negatives are all of interest, since both the false positives and false negatives still have a probability of being a pulsar. It is, however, important to minimize the amount of false negatives, since the cost of misclassifying a pulsar is larger than misclassifying a non-pulsar. This ties to the fact that all positive predictions are double checked after prediction. The true negatives are of the least amount of interest. Eight metrics are included and are defined in Table \ref{table=equations} \cite{tharwat2020classification, powers2020evaluation}.

Accuracy is a metric that measures the proportion of correctly classified samples when compared to the known labels. The recall, also known as sensitivity, is a measure of the proportion of positive cases being identified as positive. Specificity is equivalent to recall, but for the negative case. Precision measures how many of the positively identified samples were in fact actual positives and the negative prediction value (NPV) does the same for the negative predictions. The higher both the recall and precision are, the better the model is at predicting the positive class. Three more metrics are included that are commonly used: balanced accuracy, geometric mean, and informedness. Balanced accuracy is the average of recall and specificity, supplying a way to move away from a possibly inflated accuracy value if imbalanced testing sets lead to biased prediction. Geometric mean also compensates for imbalanced data, but the focus here is placed on the training set specifically. Our training sets will be balanced, but we include geometric mean for completeness. Finally informedness provides another measure of how well the model predicts both the positive and negative classes. The best model would be the one that maximizes performance targeted at the positive class (recall and precision), while avoiding as many false negatives as possible (NPV).

\subsection{\label{sec:QSVM} Quantum Kernel Assisted Support Vector Machine}

Support vector machines are designed as maximal margin classifiers where the data is classified into two classes by a linear decision boundary \cite{noble2006support}, after calculating a kernel function value, which is simply a chosen distance measure between two feature vectors. Depending on the dimension of the problem, this separating boundary is usually a separating hyperplane and is found by specifying two parameters: the normal vector of the plane and a scalar number, which together determine the offset of the plane. The margin is the distance between the closest two data points from opposing classes and the support vectors are the most important data points used when determining the hyperplane as they are the closest points to the plane and have the highest impact on how the plane should be oriented. This assumes that the data is linearly separable in the $n$-dimensional input space. If linear separability is impossible, then a feature map $\phi (\vec{x})$ can be used to map or embed the feature vectors in input space $\bm{X}$, into a higher-dimensional feature space $\bm{F}$ where it may become linearly separable. In feature space, the kernel function is typically an inner product of two mapped feature vectors and because of the kernel trick \cite{boser1992training, scholkopf1999advances}, only the kernel function needs be evaluated, instead of processing the individual vectors in feature space.

The quantum version of SVMs uses data embedding to encode the data features into quantum mechanical states that are simulable on quantum devices \cite{havlivcek2019supervised, schuld2019quantum, hubregtsen2022training}. In this way the feature map maps the data into a feature space that is similar to Hilbert space, where the Hilbert-Schmidt inner product can be used \cite{schuld2021supervised}. This implies that the feature map in this case, maps real data to quantum mechanical states described by complex density matrices. The new feature map can be described as \cite{schuld2021supervised}:

\begin{equation}
    \phi(\vec{x}) = \ket{S(\vec{x})} \bra{S(\vec{x})} = \rho(\vec{x})
\end{equation}

\noindent where the feature map $\phi (\vec{x})$ is the outer product of a data encoded feature vector $S(\vec{x})$ with itself. Quantum kernel methods, of which quantum support vector machines are an example, can be split into two steps: data encoding, which can be done in a multitude of ways \cite{schuld2021machine} (amplitude encoding, angle embedding etc.) and measurement, which can be understood as a projective operation using some observable matrix $\hat{O}$. 

Quantum kernel methods use the inner product in feature space, as a distance measure between two feature vectors, to train the model based on these values by finding a linear decision boundary that separates two classes from one another. Stated more explicitly, training here means that the optimal measurement must be found as it determines the decision boundary. The inner product of quantum embedded states is the mutual overlap. This similarity measure is now called the quantum kernel. A matrix of pairwise inner product values can be calculated by calculating the kernel value for each combination of data points. The pairwise inner product matrix calculated for two sets of identical feature vectors results in a matrix called the kernel gram matrix. The kernel matrix in this case is calculated in feature space, meaning that quantum kernels are functions of the form:

\begin{equation}
K(\vec{x}, \vec{x}') = |\bra{\phi(\vec{x})} \hat{O} \ket{\phi(\vec{x}')}|^{2} = \text{tr} \left[ \rho(\vec{x}) \hat{O} \rho(\vec{x}') \right] ,
\end{equation}

\noindent where $\vec{x}$ and $\vec{x}'$ are two separate $n$-dimensional input vectors, each with $n$ features, $\phi$ is the feature map determined by the data embedding and $\hat{O}$ is any observable that can be omitted if the only the overlap is to be calculated. The resultant kernel function is then used for fitting a classical SVM to the kernel gram matrix for training.

The quantum kernel is used in place of a classical kernel function when the classical SVM is fitted to the training data. This implies that only the kernel matrix, or rather its elements, are calculated using the quantum device. The kernel only depends on the data encoding, which implies that the data encoding defines the minima used in training \cite{schuld2021supervised}. We will call this method a QSVM from now on.

A major drawback to QSVMs is the fact that in order to calculate the full kernel matrix, a quadratically scalable number of entries must be calculated \cite{schuld2021supervised}. One entry for every pair of input feature vectors exists. The hope is that a classically intractable quantum kernel can be found that will improve classification accuracy when compared to classical kernel functions. There is also the hope that an improvement with respect to the runtime can be found, but this might prove difficult with how long quantum circuit evaluations generally take. It is also an added benefit that SVMs generally do not suffer from barren plateaus \cite{schuld2021supervised}.

\subsection{\label{sec:QCNN}Quantum Convolutional Neural Networks}

Convolutional neural networks (CNN) are commonly used architectures for image-related machine learning tasks, such as image recognition, image segmentation, and object detection \cite{krizhevsky2012imagenet, gu2018recent}. Generic CNN architectures consist of interconnected layers, including input, output, and hidden layers, with activation functions determining neuron activation. This is arranged in a graph-like configuration where the edges between nodes indicate connection weights that determine how strongly certain nodes are linked. Training involves passing data through the network, calculating a cost function, and using gradient descent to update weights in back propagation during each epoch or iteration. CNNs have three key components: convolutional layers, pooling layers, and fully connected layers. The fully connected layers operate as a standard neural network. Convolutional layers perform matrix multiplication to learn spatial patterns and pooling layers reduce spatial dimensions to reduce computational costs.

While there are many proposals for the quantum analogue of convolutional neural networks \cite{hur2022quantum, henderson2020quanvolutional, wei2022quantum}, we focus on the framework proposed by Cong et al. \cite{cong2019quantum} and follow the design methodology of Lourens et al. \cite{lourens2023hierarchical}. Here the QCNN implements analogous convolution and pooling operations in a quantum circuit setting. The key components being weight sharing, sequential reduction of system size via pooling, and translational invariance in convolution steps. These operations are applied on a circuit architectural level, where a convolution consists of unitary operations (also called mappings) applied to all available qubits in a given layer. Identical unitaries are utilised, enabling translational invariance and the sharing of weights. The choice of unitary is arbitrary, but should be chosen in such a way to minimize circuit-based noise, while retaining the ability to learn patterns in the data.

Pooling consists of using a portion of the available qubits as control qubits and applying controlled rotations on the targets. This leads to a reduction in system size and allows the number of parameters to scale logarithmically with circuit depth. Pooling introduces non-linearity to the model while also reducing its computational overhead and is analogous to coarse-graining in the classical case. Information is lost, but by connecting two qubits through a separate unitary mapping, in our case a simple CNOT gate, some information is retained for use in classification. 

Convolution and pooling operations are repeated until the system size is sufficiently small, in our case when there is one solitary qubit remaining. Circuit architecture may be changed to find possible relationships in the data, such as changing the solitary qubit. We show an example QCNN in Figure \ref{fig=QCNN Circuit} along with it's corresponding directed graph representation \cite{lourens2023hierarchical}. In the digraph representation, nodes correspond to qubits and edges to unitaries applied between them. This representation is useful for defining hyperparameters such as strides, steps and offsets that are used to optimise circuit architecture. One advantage of the QCNN is its flexibility in choice of architecture and Lourens et al. \cite{lourens2023hierarchical} showed how changing the layout of unitaries on the circuit can greatly improve model performance. Finally, all variational circuits require a data encoding step, just like for kernel methods. The choice of data encoding is really important for the effectiveness of variational quantum circuits \cite{schuld2021effect}.

\section{\label{sec:Methodology} Methodology \protect}

\subsection{\label{sec:data set}The HTRU-2 data set and Feature Selection}

The HTRU-2 data set used for this comparitive study has already been formatted for use in machine learning and consists of 8 features or inputs per candidate sample with a singular classification label of either 0 or 1, where 1 is the label of the positive class (a pulsar). There are eight features that can be grouped into two groups of four, where each of the features are statistical elements of a candidate's integrated field and DM-SNR curves. The four statistical elements of these two properties are the mean, standard deviation, excess kurtosis, and skewness. The data set contains 16259 pulsar candidates, where 1639 have been confirmed to be pulsars manually by experts. To read more about what the integrated field and DM-SNR curves are and why they are important for pulsar classification see \cite{lyon2016pulsars, misc_htru2_372}.

\begin{figure}[t]
 \begin{center}
   \includegraphics[width=\columnwidth ,height=0.754\columnwidth]{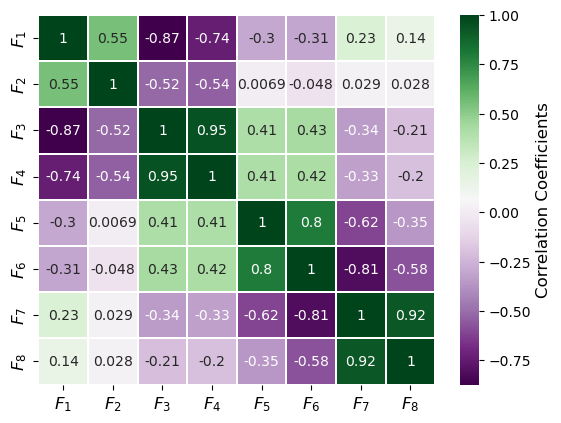}
  \caption{\label{fig=corr} A correlation matrix showing the correlation coefficients between each possible pair of the features in the HTRU-2 data set.}
  \centering
 \end{center}
\end{figure}

The features were all normalized to be between $0$ and $\pi$ radians. This is done because of the rotational encoding strategy chosen. The full data set was split in two at a ratio of 70:30. A specified number of samples from $70\%$ of the data set, in this case a balanced number of positive and negative class candidates, were sampled randomly and used as training sets. Another set of samples was also created by randomly sampling from the remaining $30\%$. This set was used as the testing set. This allows for creating training and testing sets of any size if necessary.

A specific case of batching the training set for the QCNN was done as follows: from the balanced training sets already created through random sampling, more random sampling was used to create balanced batches of $10$ candidates each that were used in each epoch. The way in which this was done, allowed for sample overlap, meaning some batches might have duplicate entries. 

\subsection{\label{sec:Main Our QSVM} QSVM Circuit}

We kept the quantum circuit depth as low as possible to circumvent decoherence noise. The quantum circuit used is an extension of the circuit used in \cite{Schuld_2021_demo}. The features were embedded as the phase arguments of Y-gates applied to each of the 8 qubits, one for each feature. This is called angle embedding and is mathematically represented by applying a Y-rotation gate to the initial state $\ket{0}$, where the parameter specifying by how much the rotation should be applied is a specific feature from a feature vector. The rotation operator can be written as $R_Y(x_i) = e^{-i \frac{x_i}{2}\sigma_Y}$, where $\sigma_Y$ is one of the Pauli matrices.

\begin{equation}
    R_Y(x_i)\ket{0},
\end{equation}

\noindent where $R_Y(x_i)$ is the Y-rotation gate that is applied to one of the 8 qubits. Each qubit was rotated by a Y-rotation of $x_i$, where the argument was the $i^{th}$ entry to a feature vector. Keeping to the simplistic design a second set of gates, the complex conjugate of the angle embedding this time only with another set of 8 features from a separate feature vector, was applied to the 8 qubits.

\begin{equation}
    R_Y^{\dagger}(x_i')R_Y(x_i)\ket{0} = e^{-i(\frac{x_i - x_i'}{2}) \sigma_Y}\ket{0},
\end{equation}

\noindent where we now have a term, $x_i - x_i'$, referencing the difference between two features from two separate feature vectors. This difference forms part of the distance measure required by SVMs. The combined effect of the two sets of Y-rotation gates can be written as unitary operators $S(\vec{x})$ and $S^{\dagger}(\vec{x}')$ and their operation on the 8-qubit initial state can be written as:

\begin{equation}
    S^{\dagger}(\vec{x}')S(\vec{x})\ket{00\ldots0},
\end{equation}

\noindent where $\ket{00\ldots0}$ is the 8-qubit initial state vector. Afterwards a projective measurement was taken by projecting the state down to the 8-qubit initial state's density matrix, $\rho = \ket{00\ldots0} \bra{00\ldots0}$. In other words the density matrix serves as part of a projector matrix before measurements are done in the computational basis. This step forms the final part of calculating the inner product. Without it the measurement would only produce the state after two rotations. This has the effect of calculating the inner product in the computational basis without performing a SWAP test \cite{Schuld_2021_demo} and becomes the kernel function that will be used along with fitting a classical support vector machine to the training data for prediction. This can be seen in the following:

\begin{equation}
    \bra{00\ldots0} S(\vec{x}')S^{\dagger}(\vec{x})\hat{O} S^{\dagger}(\vec{x}')S(\vec{x})\ket{00\ldots0},
\end{equation}

\noindent where $\bra{00\ldots0} S(\vec{x}')S^{\dagger}(\vec{x})\hat{O}$ is the projector used and a final expression can be found:

\begin{equation}
    |\bra{00\ldots0} S^{\dagger}(\vec{x}')S(\vec{x})\ket{00\ldots0}|^2,
\end{equation}

\begin{figure}[t]
 \centering
   \includegraphics[width=0.550\columnwidth, height=0.733\columnwidth]{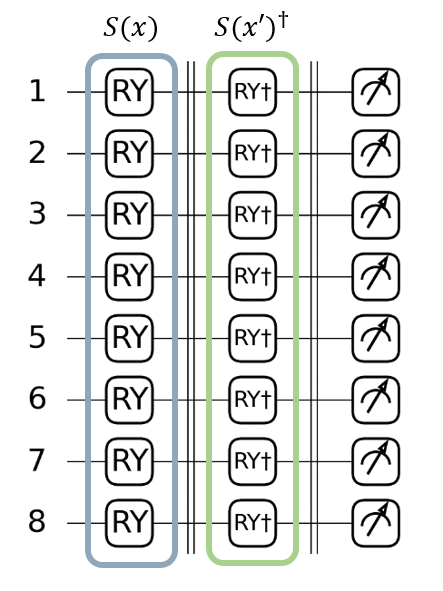}
  \caption{\label{fig=QSVM circuit} The QSVM circuit. We simply apply angle embedding (rotational encoding) using $Y$-rotation gates, followed by applying the complex conjugate transpose of this embedding using features from a separate feature vector as parameters. Measurement is done in the computational basis only after a projector matrix was applied to help calculate the inner product without a SWAP test \cite{Schuld_2021_demo}.}
\end{figure}

\noindent that is just the kernel discussed before. The above was done for the entire training set. The consequence of how kernel values are calculated (one value exists for each possible pair of features from two separate feature vectors) the scaling of this method is quadratic. This implies that the kernel will become classical intractable at some point. Therefore, since the kernel function is hard to simulate classically, it is the hope that an actual quantum device would be able to calculate it and reap the benefits of it if there are any. We estimate the training time of the training step of the algorithm on a real device by measuring the time it takes for one circuit execution to run and then multiplying this runtime with the total number of circuit executions. This would extrapolate the training time for more circuit executions. From our perspective, runtime is the time it takes the model to train on the training set. The predicting time is not irrelevant though, as for QSVMs it scales with the size of the training and testing set. In other words to estimate the training time we use:

\begin{equation}
    \Delta t_{\text{Training}} = n_{\text{CE}} \times \Delta t_{\text{CE}},
    \label{equation=runtime}
\end{equation}

\noindent where $\Delta t_{\text{Training}}$ is the total training time, $n_{\text{CE}}$ the number of circuit executions and $\Delta t_{\text{CE}}$ is the training time of one circuit execution. The total number of circuit executions can be calculated using:

\begin{equation}
    n_{\text{CE}} = n_{\text{train}}^2,
\end{equation}

\noindent where $n_{\text{train}}$ is the number of training samples. The prediction time can similarly be calculated, but instead the number of circuit executions is calculated using the number of testing samples:

\begin{equation}
    n_{\text{CE}} = n_{\text{train}} \times n_{\text{test}}.
\end{equation}

No attempt was made to batch the training data and to implement incremental learning, as kernel-assisted support vector machines are one shot solvers of a maximal margin classifier type that fits the decision boundary to training data in one go, meaning no iterations are required. The quantum circuit diagram used as explained, is illustrated in Figure \ref{fig=QSVM circuit}. This was all done using the Python libraries Pennylane and scikit-learn.

\subsection{\label{sec:Our QCNN} QCNN Circuit}

Quantum convolutional neural network architecture and consequently the architechture of the quantum circuit representing the QCNN are influenced by the shape of the convolutional and pooling steps. Our circuit makes use of eight qubits, three convolutional steps and three pooling steps, before measuring the last remaining qubit. A stride of five was chosen for the convolutional steps, meaning qubits 1 and 6 will be connected by an edge in the first convolutional step. This will have an effect throughout the circuit. We choose two-qubit unitaries for the convolution steps and attempt to maintain a short circuit depth in our experiments. Convolutional steps have the effect of cross-correlating the two qubits the operator is applied to, which also results in the sharing of weights/parameters in a neural network representation of the circuit. The chosen unitary can be seen in equation \ref{equation=unitary}.

\begin{figure*}[t]
 \begin{center}
   \includegraphics[width=\textwidth ,height=0.404\textwidth]{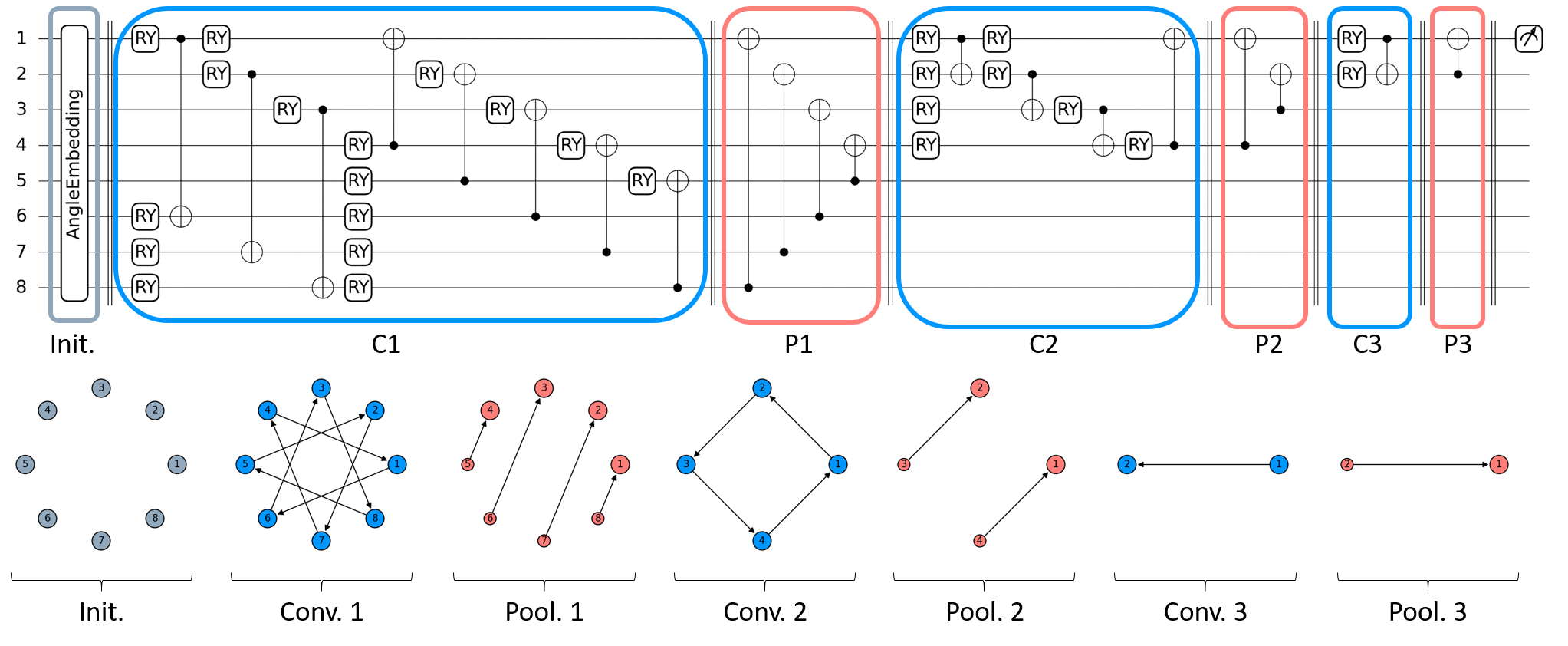}
  \caption{\label{fig=QCNN Circuit} Top: The QCNN circuit. Note the stride of $5$ in the convolutional steps. The rotational encoding, convolution, and pooling steps are all illustrated as a collection of gates inside coloured blocks. Bottom: The directed graph representation of the entire circuit. Edges in the the convolutional steps are the unitary specified earlier and edges in the pooling steps are CNOT gates. Directed edge arrows indicate which node (qubit) is the target qubit.}
  \centering
 \end{center}
\end{figure*}

\begin{equation}
    U | q_1 \rangle \otimes | q_2 \rangle = \text{CNOT} \cdot R_{Y1}(\theta_1) \cdot R_{Y2}(\theta_2) | q_1 \rangle \otimes | q_2 \rangle,
    \label{equation=unitary}
\end{equation}

\noindent where the $R_{Y1}$ and $R_{Y2}$ are y-rotation gates applied to each qubit and the rotation parameters $\theta_1$ and $\theta_2$ will be optimised every iteration. After the convolutional steps, pooling steps reduce the number of qubits to an eventual solitary qubit that is to be measured for output. In the pooling steps we use CNOT gates to represent the directed edges. Figure \ref{fig=QCNN Circuit} shows an example QCNN circuit with the directed graph representation

We considered two cases: 1 - using the entire training set for training every epoch and 2 - batching the training data into smaller batches of 10 samples or candidates per batch for training every epoch. The way the batching was done meant that there was chance of sample overlap between batches, with this overlap being almost guaranteed if the batch size of 10 multiplied by the number of epochs was comparable to the training set size. Ideally it would be best to use large training sets. In both cases the learning rate was set to 0.01 and the iteration limit was set to 150 epochs. The chosen loss function to minimize is the binary cross-entropy loss:

\begin{equation}
\begin{aligned}
\text{BCEL}(y_{\text{pred}}, y_{\text{test}}) = -\Bigl(y_{\text{test}} \cdot \log(y_{\text{pred}}) + \\
(1 - y_{\text{test}}) \cdot \log(1 - y_{\text{pred}})\Bigr).
\end{aligned}
\end{equation}

The QCNN quantum circuit used is illustrated in Figure \ref{fig=QCNN Circuit}. The Python libraries PyTorch and Pennylane were both used in conjunction to iterate over the forward pass, back pass and gradient updating steps.

A consequence of both the circuit architecture and how the method is implemented is a different scaling to the QSVM. We can still approximate the total training time by using equation \ref{equation=runtime}, but now the total circuit executions can be easily controlled and calculated using the total number of epochs and samples used:

\begin{equation}
    n_{\text{CE}} = n_{\text{Epochs}} \times n_{\text{Train}},
\end{equation}

\noindent where the number of epochs and training samples can be controlled. Estimating the prediction time is much simpler:

\begin{equation}
    n_{\text{CE}} = n_{\text{Test}},
\end{equation}

since only one circuit execution is performed per candidate.

\section{\label{sec:Results} Results \protect}

\subsection{\label{sec:Simulations}Simulations}

We simulate the above circuits, using Pennylane, and plot the confusion matrices. This gives us an indication of how the methods compare with respect to performance on the chosen data set. The classical device used to simulate all the noise-free and noisy simulations has the following specifications: Lenovo Laptop Intel(R) Core(TM) i7-6700HQ CPU @ 2.60GHz 2.60 GHz, with 16,0 GB (15,8 GB usable) RAM and a GTX 960M GPU, which was not used to supplement the computational power.

\begin{figure}[t]
 \begin{center}
   \includegraphics[width=\columnwidth ,height=1.652\columnwidth]{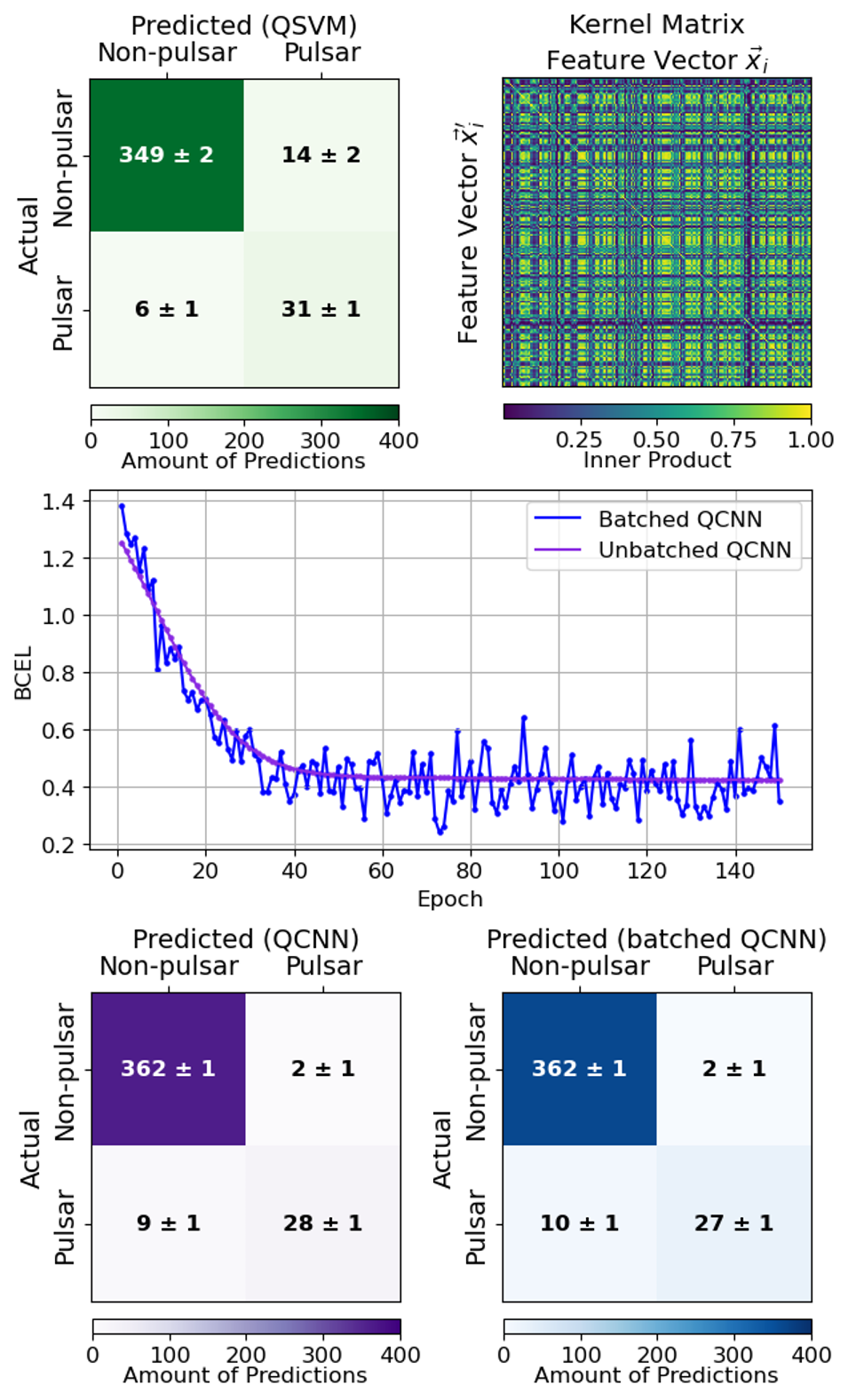}
  \caption{\label{fig=cf matrices} Confusion matrices for runs at $200$ training samples and $400$ testing samples using QSVMs (Green), QCNNs (Purple) and batched QCNNs (Blue). The confusion matrices of both QCNN approaches are nearly identical, which is an indication that batching should be strongly considered. The two loss curves are each an example of one of the convergent loss optimization processes. An example of a kernel gram matrix (top right) is provided for $200$ training samples. Each entry in the confusion matrices is the average of six distinct runs and the uncertainty is the standard error.}
  \centering
 \end{center}
\end{figure}

The confusion matrices from Figure \ref{fig=cf matrices} (which are plotted for runs at 200 training samples and 400 testing samples) show that both methods perform well at classifying pulsars correctly. Both methods are good at correctly predicting negative cases. False positives are harder to avoid for the QSVM than for the QCNN and the reverse is true for the false negatives. The QSVM predicted a higher proportion of the positive cases correctly, however, the QCNN still managed to correctly predict most positive cases correctly as well. The implication of the confusion matrices is that if the QSVM model predicts a candidate as a non-pulsar, it is almost certainly a non-pulsar, however when the model predicts that a candidate is a pulsar, then that candidate might have to be double checked. The reverse applies to the QCNN. The QCNN performs worse at avoiding false negatives, which is not ideal as we want to minimise the amount of mistakes when predicting the positive case.

It is fortunate that there are more non-pulsar candidates than pulsar candidates that are correctly predicted in both cases. This causes a reduced amount of manual confirmation required, as negative predictions will never be doubled checked. This is because they form the majority of all predictions, which means there are too many to manually confirm. Even though the amount of manual labour is reduced, to be entirely sure of the positive predictions, all positive predictions will have to be manually checked, especially when applying the model to unseen data.

\begin{figure*}[t]
 \begin{center}
 \includegraphics[width=\textwidth ,height=0.814\textwidth]{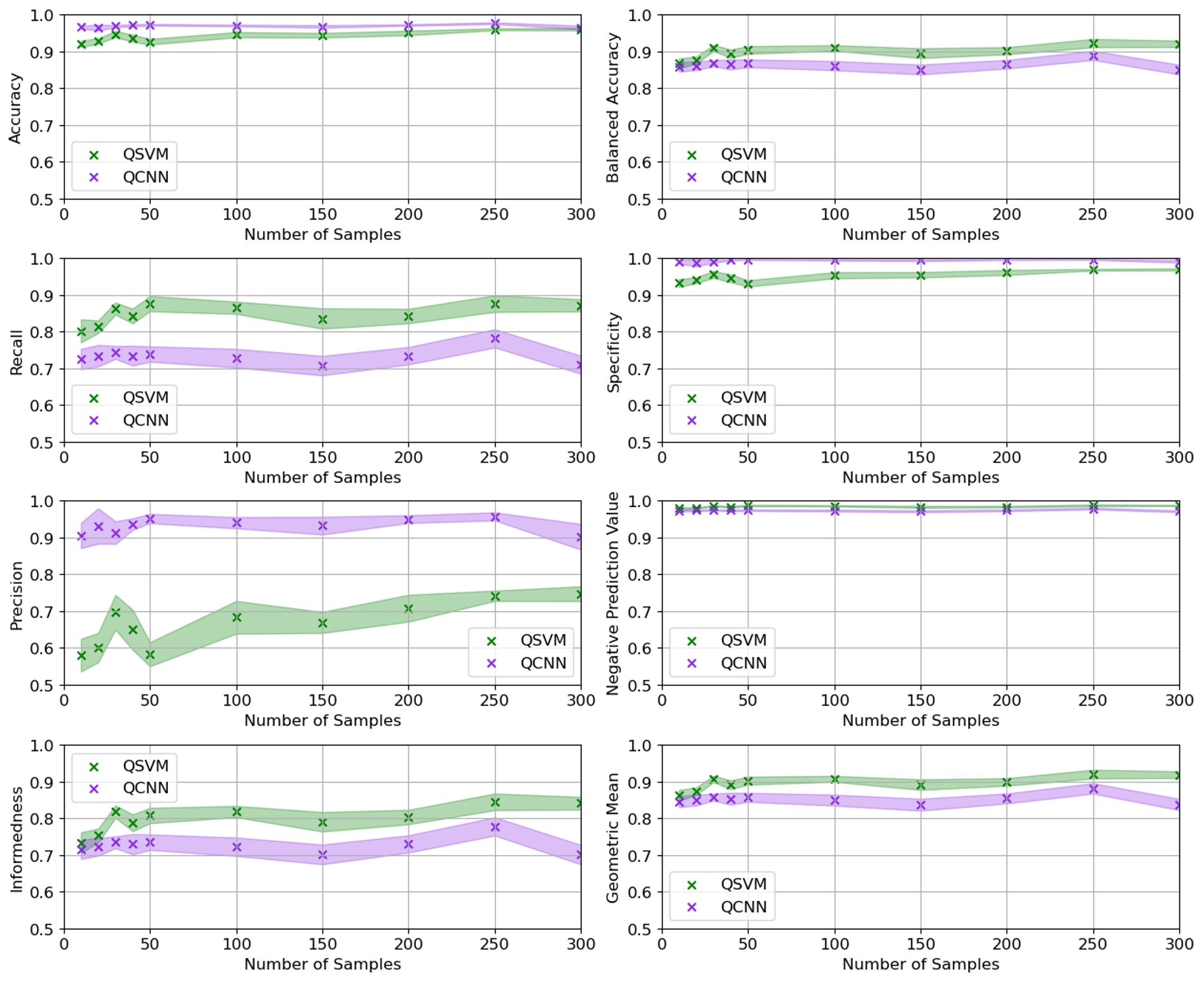}
  \caption{\label{fig=sample_test} Eight performance metrics plotted versus an increasing training sample size. The plots included (alternating from left to right going downwards) are accuracy, balanced accuracy, recall, specificity, precision, negative prediction value, informedness, and geometric mean. Averages are illustrated using x-markers and the uncertainty (standard error) is indicated as a shaded area. Each data point is the average of twelve distinct runs. Green indicates the QSVM and purple the QCNN.}
  \centering
 \end{center}
\end{figure*}

We also simulate the circuits for an increasing number of training samples. The resultant curves of eight performance metrics can be plotted as seen in Figure \ref{fig=sample_test}. The expected behaviour would be for the performance metrics to increase as sample size is increased, as more samples would typically lead to a more optimized model. This is marginally true for the lower sample sizes (less than 50 samples) where a slight increase can be observed. It is also observed in the QSVM precision plot and in isolated cases for informedness, geometric mean, balanced accuracy and recall. The near constant behavior throughout the QCNN case indicates that the saturation point for training is reached early and a possible explanation for this is training over many epochs using the entire training set. 

The uncertainty at lower samples is also more prominent, which implies that for a limited number of samples, training is not satisfactory. This is observed in the accuracy, specificity, and precision curves. We observe that as the number of samples are increased, the uncertainty in the accuracy and specificity decreases. This indicates an improved prediction aimed towards the negative case. There is also an increase in the precision curve for the QSVM, implying more samples result in a improved prediction performance towards the positive cases. Recall and precision have a higher uncertainty and lower values than specificity, which again illustrates that it is easier to predict negative cases correctly. All positive predictions have a higher probability to be incorrect and will thus have to be manually checked. This is consistent with the results from the confusion matrices.

Two important points worth noting, are the uncertainty and fluctuating behaviour in the balanced accuracy, geometric mean, and informedness. All three metrics depend on recall as a metric, which indicates the performance on positive cases. Since there are more mistakes for the positive cases, and hence more uncertainty, these values tend to fluctuate more.

Finally, from a general perspective, it is clear that the models perform well when measuring in accuracy, specificity and negative prediction value as metrics and slightly worse when considering precision and recall. It depends on the sample size for the other metrics, but we note that the QSVM outperforms the QCNN on most metrics, which is a strong motivator for its use. If precision is prioritised, then the QCNN can be considered as well, since it has a precision that is high compared with the QSVM.

It is also interesting to note that changing the QCNNs architecture, for example the choice of unitary operators used, may give insight as to how certain data features might be linked. In other words QCNNs might be more interpretable than QSVMs. This is especially true for changing the striding between qubits in the convolutional step and changing the final remaining qubit after all the pooling steps. It might supply a way of gauging which features are more important or more closely related without performing PSA or feature correlation tests. An example of this is a sudden increase in accuracy when using a stride of 5, which can be attributed to an unknown correlation in the features.

\subsection{\label{sec:QCNN: Unbatched versus Batched Data}QCNN: Unbatched versus Batched Data}

We considered the effect of data batching for the QCNN, which involved randomly batching the training data into balanced groups of 10 each. This means that in each epoch of the training and optimization step, the model is training with a new set of 10 samples. The two cases can be compared in Figure \ref{fig=cf matrices}.

For the batched case it is clear that the minimization of the loss fluctuated a lot more, since it trained over less samples in each iteration. The overall result in the confusion matrices was nearly identical.

\begin{table*}[t]
\centering
\renewcommand{\arraystretch}{1.2}
\setlength{\tabcolsep}{4pt}
\begin{tabular}{ lrrcrr } 
\toprule
\textbf{Metric} & (1) \textbf{QSVM} & (2) \textbf{QCNN} & (3) \textbf{Batched QCNN} & (4)\textbf{CSVM} & (5) \textbf{CCNN} \\ 
\midrule
\text{Accuracy} & $0.950 \pm 0.006$ & $\mathbf{0.972 \pm 0.002}$ & $0.971 \pm 0.003$ & $0.972 \pm 0.004$ & $0.942 \pm 0.009$ \\ 
\text{Balanced Accuracy} & $0.902 \pm 0.009$ & $0.865 \pm 0.011$ & $0.863 \pm 0.014$ & $\mathbf{0.944 \pm 0.007}$ & $0.935 \pm 0.013$ \\
\text{Recall} & $0.842 \pm 0.018$ & $0.734 \pm 0.022$ & $0.730 \pm 0.028$ & $0.910 \pm 0.012$ & $\mathbf{0.928 \pm 0.027}$ \\
\text{Specificity} & $0.961 \pm 0.006$ & $\mathbf{0.996 \pm 0.001}$ & $\mathbf{0.996 \pm 0.001}$ & $0.978 \pm 0.004$ & $0.943 \pm 0.010$ \\
\text{Precision} & $0.707 \pm 0.033$ & $\mathbf{0.950 \pm 0.010}$ & $0.948 \pm 0.018$ & $0.817 \pm 0.030$ & $0.638 \pm 0.036$ \\
\text{Negative Prediction Value} & $0.984 \pm 0.002$ & $0.974 \pm 0.002$ & $0.973 \pm 0.003$ & $0.991 \pm 0.001$ & $\mathbf{0.992 \pm 0.003}$ \\
\text{G-Mean} & $0.899 \pm 0.009$ & $0.854 \pm 0.012$ & $0.852 \pm 0.017$ & $\mathbf{0.943 \pm 0.007}$ & $0.935 \pm 0.013$ \\
\text{Informedness} & $0.803 \pm 0.018$ & $0.730 \pm 0.021$ & $0.726 \pm 0.028$ & $\mathbf{0.888 \pm 0.014}$ & $0.871 \pm 0.026$ \\
\text{Training Time (s)} & $1241.99 \pm 29.75$ & $1262.85 \pm 9.08$ & $61.25 \pm 0.63$ & \textbf{Negligible} & $1.77 \pm 0.05$ \\
\text{Prediction Time (s)} & $2492.15 \pm 60.53$ & $12.26 \pm 0.13$ & $12.04 \pm 0.03$ & \textbf{Negligible} & $0.14 \pm 0.01$ \\
\bottomrule
\end{tabular}
\caption{The results of both the QSVM (1) and QCNN (2) approaches, as well as the batched QCNN (3), classical SVM (4), and classical CNN (5) methods. All runs were taken for models trained at $200$ training samples and testing was performed on $400$ testing samples. Each entry is the average $\pm$ standard error. The best performing method for each metric is indicated in bold font.}
\label{table:results_corrected}
\end{table*}

The results from Table \ref{table:results_corrected} shows that the gain from training over all samples in each epoch is not worth the increase in overall training time. It is therefore recommended that data batching be used for this method as it will drastically decrease the amount of time required to train the model. It is also a good idea to increase the number of epochs to cover as many samples as possible and to limit the amount of sample overlap in each batch. Note that this only applies to the way we batched the training data and can be removed entirely if unique batches are sampled. Since it is possible to use a batched QCNN to benefit from its faster training and prediction times compared to the standard QCNN approach, the batched QCNN can also be implemented on even larger batches or data set sizes. The result should be an increase in performance and a decrease in uncertainty.

\subsection{\label{sec:Classical versus Quantum}Classical versus Quantum}

The runtime plots, that follow later in Figure \ref{fig=runtime_full}, indicate that there is an intersection in the training time somewhere close to 200 training samples for both quantum models. Assuming that training time would be comparable at this sample size, generic classical versions of a SVM and a CNN were also performed to compare these methods with their classical equivalents. The results are tabulated in Table \ref{table:results_corrected}.

Comparing CNNs and SVMs separately shows that classical SVMs perform better compared to quantum SVMs when considering every metric, however, the QSVM does perform similarly when considering specificty. This means that the QSVM is up to par with the CSVM when predicting negative cases. The QCNN is also outperformed by its classical equivalent, but the difference is that its specificity is actually higher than its classical counterpart. This implies that if you wanted to solve the current problem with a quantum approach, the QCNN would be the better choice of the two, not only because it has the higher specificity, but also its faster training time. CCNNs are typically used for spatial analysis in image data, which can cause shortcomings in their performance when identifying feature vectors. This is clearly observed in the low precision value in the CCNN column. It might be possible to improve the classical CNN and achieve improved results (specifically to increase the specificity and the precision), however, it is promising that if quantum computers become more error robust and have faster operational and readout times that there are some improvements that can be made for quantum approaches. Perhaps when we transcend the NISQ era, meaning when the runtimes are reduced and noise limited enough for these methods to become viable when compared to classical methods, quantum approaches may become mainstream. 

\subsection{\label{sec:Real Devices}Real Devices and Noisy Simulations}

We compare the runtime performance of both models on real quantum devices. This is made possible by Qiskit Runtime \cite{qiskit_runtime, Qiskit}, which is a cloud-based hybrid quantum-classical computational system to implement quantum calculations and to optimize classically. Qiskit Runtime can be interfaced with Pennylane's qiskit-pennylane plugin \cite{bergholm2018pennylane}. It was observed that with both methods, in order to perform a comparative run at 200 training samples and all other parameters kept exactly as it was set in the simulations, the overall training time would be completely impractical on real devices. The QCNN and QSVM on real devices had an approximate training time of $142.00 \pm 5.10 s$ and $3.33 \pm 0.47 s$ respectively. Using the scalability statements made earlier, these training times can be used to approximate a total training time value for each method by simply multiplying it with the total number of circuit executions required to finish the entire training and predicting process. The result is that it would take longer than a day for the QCNN and QSVM approaches just to be trained. This becomes worse if we include real device prediction and queue time. These times are calculated for a reduced number of epochs of 10 in the QCNN case. Both numbers are impractical, and even if a data batching workaround was used along with reducing the number of iterations even further, the time of one circuit execution is still too long compared with the near instantaneous runtime of one iteration in the classical counterparts. The QSVM would be the better choice here, since its circuit execution time is faster, however, the scalability of the QSVM has to be factored in.

To compare the impact of noise on both circuits, focus was placed on noisy simulations. We wanted to find the relationship between the depth of the quantum circuits and their intrinsic noise. Apart from how the circuits are changed during the compilation process, where it is translated from the designed architecture to what is has to be on the quantum device due to the device's restricted basis gate set, another source of error is gate-based error. This is where any gate applied in the circuit has a probability of failing due to many factors such as: short coherence times that lead to eventual decoherence of the state, crosstalk between qubits, and imprecise control of the qubits when applying operator gates. The more gates there are in succession, the higher the impact will be. Device specific qubit connectivity can lead to an overall greater depth during compilation which would accentuate these problems. This happens because certain multi-qubit gates require two qubits to be connected in order to be applied. If they are not connected, the compilation process tries to find another gate combination that would deliver the same result. This increases the depth, adding more gates that have probabilities to fail. 

This is the main source of error in larger depth circuits and we model the gate-based error of a quantum computer with simple bit flips, then it can be clearly seen, in Table \ref{table:noise}, that the longer depth of the QCNN circuit induces a more error prone method and the eventual measurement results will be more random. This serves as an indication that you would rather choose the QSVM, since it is more resilient to noise during the current NISQ era. We increase the probability of a single gate undergoing a bit flip to see just how random the results can get. This probability is the analogue of the median error rates that can be found on the device page on the IBM Quantum Experience page \cite{IBMQuantumPlatform}.

\begin{table}[b]
\centering
\renewcommand{\arraystretch}{1.2}
\setlength{\tabcolsep}{6pt}
\begin{tabular}{ lrrr } 
\toprule
\textbf{Noise Level} & (1) \text{QSVM} & (2) \text{QCNN} \\
\midrule
\textbf{p = 0} & $0.911 \pm 0.023$ & $0.914 \pm 0.026$ \\
\textbf{p = 0.01} & $0.879 \pm 0.021$ & $0.852 \pm 0.03$ \\
\textbf{p = 0.1} & $0.836 \pm 0.031$ & $0.758 \pm 0.055$ \\
\textbf{p = 0.5} & $0.501 \pm 0.018$ & $0.500 \pm 0.000$ \\
\bottomrule
\end{tabular}
\caption{The effect of circuit-based noise due to the depth of the quantum circuits on method performance indicated using balanced accuracy. A decreasing trend is observed for both methods. The QSVM case performs better at increasing noise probabilities than the QCNN case because of larger depth in the QCNN circuit. Both methods yield random results at $p = 0.5$. Each table entry is the average of 6 distinct runs and the uncertainties are the standard deviations.}
\label{table:noise}
\end{table}

As the probability of gate-based bit flips are increased for both the QSVM and QCNN we see a clear downward trend in the balanced accuracy. This is accentuated for the QCNN case where the falloff is more rapid. The effect of noise is more pronounced for the QCNN because of its larger depth. More gates in succession have a higher probability of undergoing a bit flip. These measurements were taken for 50 training samples and 100 testing samples to reduce computational costs when simulating noisy quantum circuits. Finally for a bit flip probability of $0.5$, the measurements become truly random, which is exactly what is expected. The reason balanced accuracy was used, and none of the other metrics, was to use a metric that is not inflated when more negative cases than positive cases were predicted correctly. We used a metric not biased towards either the positive or negative cases.

\begin{figure}
 \begin{center}
   \includegraphics[width=0.7\columnwidth ,height=0.489\columnwidth]{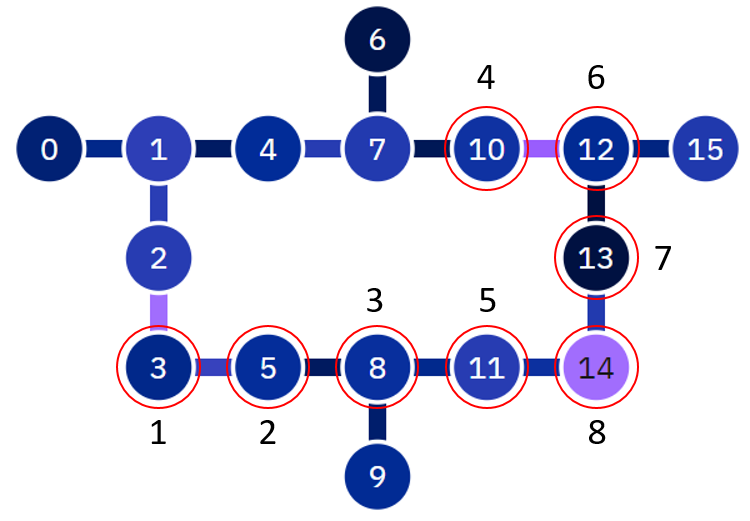}
  \caption{\label{fig=q_connec} The qubit connectivity diagram for the IBMQ Guadalupe device. The qubits marked with red circles formed part of the QCNN approach and the order in which they were assigned in the circuit is indicated by black numbers. Qubits $0$ to $7$ (white text) were used for the QSVM.}
  \centering
 \end{center}
\end{figure}

The quantum device used to approximate the total runtime of the two circuits was the IBMQ Guadalupe device, which has 16 qubits of which the ones marked in red in Figure \ref{fig=q_connec} were used for the QCNN. The black numbers indicate which qubit each node represented in the actual implementation. Qubits 0 to 7 were used for the QSVM as no qubit connectivity was explicitly required.

\subsection{\label{sec:Runtime} Runtime}

\begin{figure*}[t]
 \begin{center}
   \includegraphics[width=\textwidth ,height=0.506\textwidth]{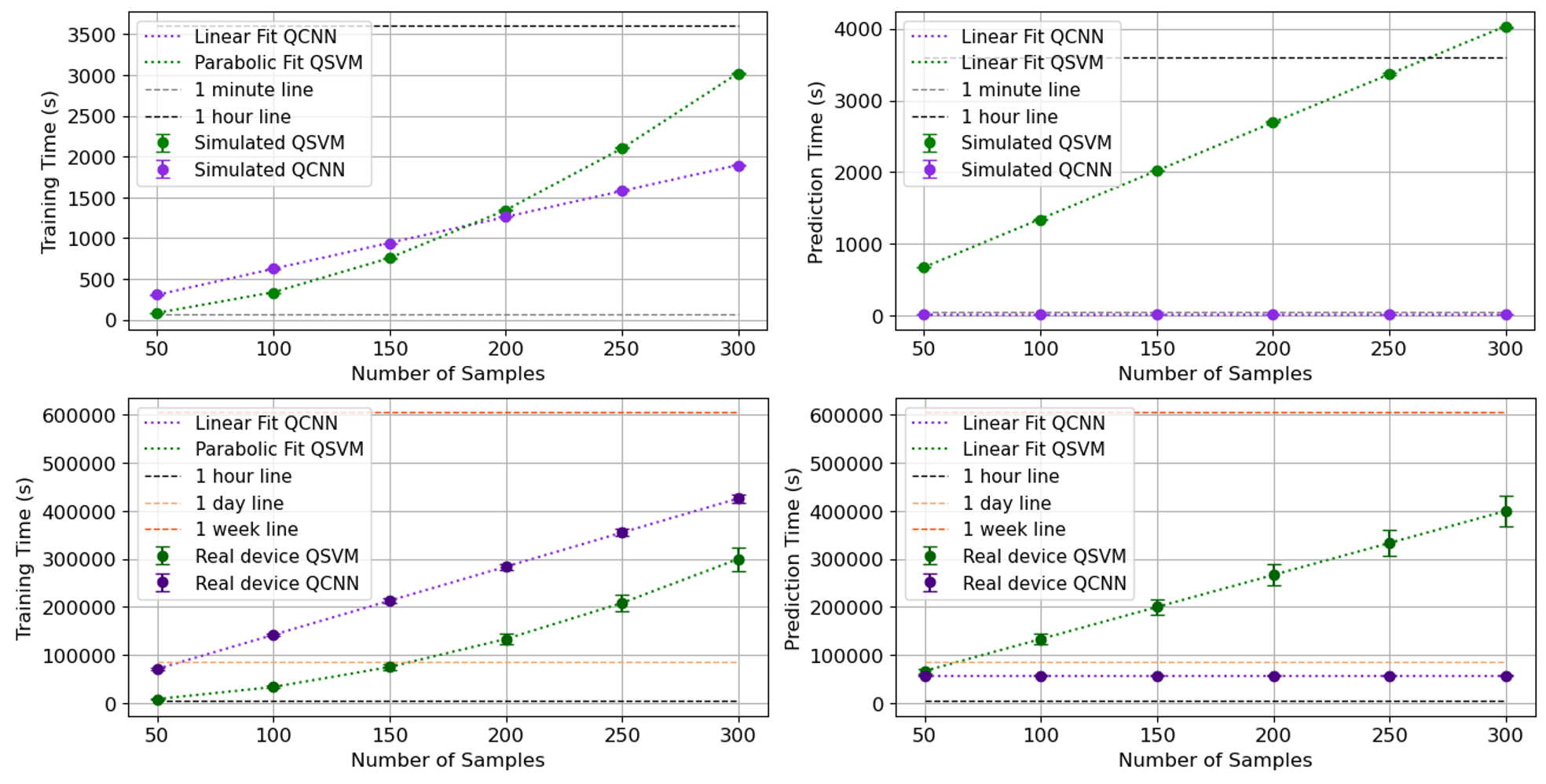}
  \caption{\label{fig=runtime_full} Training and prediction time in seconds versus an increasing number of samples using a testing set of $400$ samples for (Top) simulated runs and (Bottom) approximated real device runs. Both training plots show a linear scaling in the QCNN and a quadratic scaling in the QSVM that has been described before. The prediction time plots indicate that the QSVM takes longer to perform predictions. This is because prediction requires reference to the training set used. Simulated and real device results are taken and approximated for QCNN runs at $150$ and $10$ epochs, respectively. Note the scale of the y-axis in the real device plots.}
  \centering
 \end{center}
\end{figure*}

Another figure of merit we considered, as discussed earlier, is runtime. This includes both training and prediction time. The runtime plots can be found in Figure \ref{fig=runtime_full}.

The training time scaling of both methods can be seen from Figure \ref{fig=runtime_full}. The QSVM scales quadratically, while the QCNN scales linearly, with the intersection between the two curves indicating that for a low number of samples the QSVM is faster. The QCNN will train faster on a larger sample size. The same intersection for the real devices cannot be observed in the figure, but it is clear the two curves are tending towards a similar intersection.

The prediction time plots show that for QCNNs the prediction time remains constant compared to the linear scaling of the QSVM. This is because QCNN prediction is done for a constant $400$ samples. No reference is made to the training set as in the QSVM case. The linear scaling of the prediction time for QSVMs is because it scales linearly with the number of training samples, while the number of prediction samples are held constant. This is a significant aspect to consider when QSVMs are used, since using a QSVM on unseen data will take longer.

\section{\label{sec:Conclusion}Conclusion}

In this comparative study we implemented an 8-qubit quantum-enhanced support vector machine and an 8-qubit quantum convolutional neural network. The two methods were compared by performing noise free and noisy simulations as well as a limited real device implementation that allowed for approximation of a possible fully implemented run on real devices. The training time and prediction time for both methods were also compared for many sample sizes. Both methods show improvement over the known quantum approach \cite{kordzanganeh2021quantum} and are also lower depth, which helps with noise robustness.

QSVMs are easy to implement and have a small depth, which is really important for avoiding decoherence and noise in real device runs. QSVMs, however, scale quadratically with sample size, which makes them really unattractive for big-data problems. The kernel function used by the SVM is a black box, meaning that there is no way to tell how the algorithm relates features to prediction. They are, however, really effective at a lower number of training samples and in this regime they can definitely be considered.

QCNNs with a larger depth and a more complex circuit are more prone to noise, but does indeed provide more efficient training and prediction than from the QSVM. A specificity that improved over our classical approaches is also observed. Although the approach is not completely transparent, it also supplies a method to get an intuition as to how the different features are linked to one another. The special case of batching the training data resulted in near identical results (the same high specificity was observed here as well) which makes QCNNs viable for a larger number of samples.

It is true that both methods have their downsides and advantages, but it should be clear that the method of choice should depend on the training set size. Judging by its ease of implementation, shorter depth making it more noise resistant, guarantees of performance compared to classical methods and its shorter runtime for a low sample size when simulated and on real devices, QSVMs should be preferred comparatively over QCNNs. Larger sample sizes after the two training time curves have intersected demands that QCNNs be implemented instead, definitely making use of data batching, however the larger depth will introduce a lot of noise into the system, which means that error mitigation techniques will be required.

It should be clear that quantum machine learning can be used for astronomical object classification and real-world binary classification in general, but in the current NISQ era QML methods applied to classical data falls short when compared to classical methods.

A broad architecture search was already performed by changing some of the hyperparameters like striding, but it may be possible to find a QCNN circuit architecture that performs even better than the one considered here by using different unitaries in the convolutional or pooling steps. It would be good idea to do a more comprehensive circuit architecture search for other 8-qubit QCNN architectures that would provide possible improvements to classification accuracy. This is made easier by using the HierarQcal package, which has been used for a similar application in \cite{lourens2023hierarchical}. Circuit-based noise would be the main limiting factor and is the reason why we kept the QCNN circuit as simple as possible.

There are many ways in which this work can be extended. Multi-dimensional classification, where there are more than two classes could be an interesting extension. A typical example of multi-classes here would be standard and millisecond pulsars. These methods can also be applied to other data sets to find out if the same behaviour persists. Once better quantum computers are built, either through fault tolerance, improved qubit connectivity or faster operational and readout times, the real device findings can be revisited to confirm if they still hold true then. Applying the methods on real devices may or may not require improvement using error mitigation techniques. This would, however, be the perfect indicator of the viability of these methods for the application of pulsar classification on real devices. Trying different approaches to the problem of pulsar classification such as quantum anomaly detection may yield interesting results.  This would involve training a quantum model on only the non-pulsar data and then testing the model on a testing set consisting of a mix of pulsar and non-pulsar samples to see if the model is able to effectively pick out the anomalous pulsars.

\section*{Acknowledgments}

This work was funded by the South African Quantum Technology Initiative (SA QuTi) through the Department of Science and Innovation of South Africa.

We also wish to acknowledge Dr. Maria Schuld for clarifying theoretical points regarding support vector machines and some of the initial work done by Shivani Pillay on classification of pulsars using quantum kernel methods. 


\bibliography{main} 

\providecommand{\noopsort}[1]{}\providecommand{\singleletter}[1]{#1}%
\begin{thebibliography}{49}%
\makeatletter
\providecommand \@ifxundefined [1]{%
 \@ifx{#1\undefined}
}%
\providecommand \@ifnum [1]{%
 \ifnum #1\expandafter \@firstoftwo
 \else \expandafter \@secondoftwo
 \fi
}%
\providecommand \@ifx [1]{%
 \ifx #1\expandafter \@firstoftwo
 \else \expandafter \@secondoftwo
 \fi
}%
\providecommand \natexlab [1]{#1}%
\providecommand \enquote  [1]{``#1''}%
\providecommand \bibnamefont  [1]{#1}%
\providecommand \bibfnamefont [1]{#1}%
\providecommand \citenamefont [1]{#1}%
\providecommand \href@noop [0]{\@secondoftwo}%
\providecommand \href [0]{\begingroup \@sanitize@url \@href}%
\providecommand \@href[1]{\@@startlink{#1}\@@href}%
\providecommand \@@href[1]{\endgroup#1\@@endlink}%
\providecommand \@sanitize@url [0]{\catcode `\\12\catcode `\$12\catcode
  `\&12\catcode `\#12\catcode `\^12\catcode `\_12\catcode `\%12\relax}%
\providecommand \@@startlink[1]{}%
\providecommand \@@endlink[0]{}%
\providecommand \url  [0]{\begingroup\@sanitize@url \@url }%
\providecommand \@url [1]{\endgroup\@href {#1}{\urlprefix }}%
\providecommand \urlprefix  [0]{URL }%
\providecommand \Eprint [0]{\href }%
\providecommand \doibase [0]{https://doi.org/}%
\providecommand \selectlanguage [0]{\@gobble}%
\providecommand \bibinfo  [0]{\@secondoftwo}%
\providecommand \bibfield  [0]{\@secondoftwo}%
\providecommand \translation [1]{[#1]}%
\providecommand \BibitemOpen [0]{}%
\providecommand \bibitemStop [0]{}%
\providecommand \bibitemNoStop [0]{.\EOS\space}%
\providecommand \EOS [0]{\spacefactor3000\relax}%
\providecommand \BibitemShut  [1]{\csname bibitem#1\endcsname}%
\let\auto@bib@innerbib\@empty
\bibitem [{\citenamefont {Lyne}\ and\ \citenamefont
  {Graham-Smith}(2012)}]{lyne2012pulsar}%
  \BibitemOpen
  \bibfield  {author} {\bibinfo {author} {\bibfnamefont {A.}~\bibnamefont
  {Lyne}}\ and\ \bibinfo {author} {\bibfnamefont {F.}~\bibnamefont
  {Graham-Smith}},\ }\href@noop {} {\emph {\bibinfo {title} {Pulsar
  astronomy}}},\ \bibinfo {number} {48}\ (\bibinfo  {publisher} {Cambridge
  University Press},\ \bibinfo {year} {2012})\BibitemShut {NoStop}%
\bibitem [{\citenamefont {Kippenhahn}\ \emph {et~al.}(1990)\citenamefont
  {Kippenhahn}, \citenamefont {Weigert},\ and\ \citenamefont
  {Weiss}}]{kippenhahn1990stellar}%
  \BibitemOpen
  \bibfield  {author} {\bibinfo {author} {\bibfnamefont {R.}~\bibnamefont
  {Kippenhahn}}, \bibinfo {author} {\bibfnamefont {A.}~\bibnamefont
  {Weigert}},\ and\ \bibinfo {author} {\bibfnamefont {A.}~\bibnamefont
  {Weiss}},\ }\href@noop {} {\emph {\bibinfo {title} {Stellar structure and
  evolution}}},\ Vol.\ \bibinfo {volume} {192}\ (\bibinfo  {publisher}
  {Springer},\ \bibinfo {year} {1990})\BibitemShut {NoStop}%
\bibitem [{\citenamefont {Maoz}(2016)}]{maoz2016astrophysics}%
  \BibitemOpen
  \bibfield  {author} {\bibinfo {author} {\bibfnamefont {D.}~\bibnamefont
  {Maoz}},\ }\href@noop {} {\emph {\bibinfo {title} {Astrophysics in a
  Nutshell}}},\ Vol.~\bibinfo {volume} {16}\ (\bibinfo  {publisher} {Princeton
  university press},\ \bibinfo {year} {2016})\BibitemShut {NoStop}%
\bibitem [{\citenamefont {Manchester}\ and\ \citenamefont
  {Taylor}(1977)}]{manchester1977pulsars}%
  \BibitemOpen
  \bibfield  {author} {\bibinfo {author} {\bibfnamefont {R.~N.}\ \bibnamefont
  {Manchester}}\ and\ \bibinfo {author} {\bibfnamefont {J.~H.}\ \bibnamefont
  {Taylor}},\ }\bibfield  {title} {\bibinfo {title} {Pulsars},\ }\href@noop {}
  {\bibfield  {journal} {\bibinfo  {journal} {Irish Astronomical Journal}\ }
  (\bibinfo {year} {1977})}\BibitemShut {NoStop}%
\bibitem [{\citenamefont {Hulse}\ and\ \citenamefont
  {Taylor}(1975)}]{hulse1975discovery}%
  \BibitemOpen
  \bibfield  {author} {\bibinfo {author} {\bibfnamefont {R.~A.}\ \bibnamefont
  {Hulse}}\ and\ \bibinfo {author} {\bibfnamefont {J.~H.}\ \bibnamefont
  {Taylor}},\ }\bibfield  {title} {\bibinfo {title} {Discovery of a pulsar in a
  binary system},\ }\href@noop {} {\bibfield  {journal} {\bibinfo  {journal}
  {The Astrophysical Journal}\ }\textbf {\bibinfo {volume} {195}},\ \bibinfo
  {pages} {L51} (\bibinfo {year} {1975})}\BibitemShut {NoStop}%
\bibitem [{\citenamefont {Stairs}(2003)}]{stairs2003testing}%
  \BibitemOpen
  \bibfield  {author} {\bibinfo {author} {\bibfnamefont {I.~H.}\ \bibnamefont
  {Stairs}},\ }\bibfield  {title} {\bibinfo {title} {Testing general relativity
  with pulsar timing},\ }\href@noop {} {\bibfield  {journal} {\bibinfo
  {journal} {Living Reviews in Relativity}\ }\textbf {\bibinfo {volume} {6}},\
  \bibinfo {pages} {1} (\bibinfo {year} {2003})}\BibitemShut {NoStop}%
\bibitem [{\citenamefont {Foster~III}(1990)}]{foster1990constructing}%
  \BibitemOpen
  \bibfield  {author} {\bibinfo {author} {\bibfnamefont {R.~S.}\ \bibnamefont
  {Foster~III}},\ }\href@noop {} {\emph {\bibinfo {title} {Constructing a
  pulsar timing array}}}\ (\bibinfo  {publisher} {University of California,
  Berkeley},\ \bibinfo {year} {1990})\BibitemShut {NoStop}%
\bibitem [{\citenamefont {McLaughlin}(2013)}]{mclaughlin2013north}%
  \BibitemOpen
  \bibfield  {author} {\bibinfo {author} {\bibfnamefont {M.~A.}\ \bibnamefont
  {McLaughlin}},\ }\bibfield  {title} {\bibinfo {title} {The north american
  nanohertz observatory for gravitational waves},\ }\href@noop {} {\bibfield
  {journal} {\bibinfo  {journal} {Classical and Quantum Gravity}\ }\textbf
  {\bibinfo {volume} {30}},\ \bibinfo {pages} {224008} (\bibinfo {year}
  {2013})}\BibitemShut {NoStop}%
\bibitem [{\citenamefont {Verbiest}\ \emph {et~al.}(2016)\citenamefont
  {Verbiest}, \citenamefont {Lentati}, \citenamefont {Hobbs}, \citenamefont
  {van Haasteren}, \citenamefont {Demorest}, \citenamefont {Janssen},
  \citenamefont {Wang}, \citenamefont {Desvignes}, \citenamefont {Caballero},
  \citenamefont {Keith} \emph {et~al.}}]{verbiest2016international}%
  \BibitemOpen
  \bibfield  {author} {\bibinfo {author} {\bibfnamefont {J.}~\bibnamefont
  {Verbiest}}, \bibinfo {author} {\bibfnamefont {L.}~\bibnamefont {Lentati}},
  \bibinfo {author} {\bibfnamefont {G.}~\bibnamefont {Hobbs}}, \bibinfo
  {author} {\bibfnamefont {R.}~\bibnamefont {van Haasteren}}, \bibinfo {author}
  {\bibfnamefont {P.~B.}\ \bibnamefont {Demorest}}, \bibinfo {author}
  {\bibfnamefont {G.}~\bibnamefont {Janssen}}, \bibinfo {author} {\bibfnamefont
  {J.-B.}\ \bibnamefont {Wang}}, \bibinfo {author} {\bibfnamefont
  {G.}~\bibnamefont {Desvignes}}, \bibinfo {author} {\bibfnamefont
  {R.}~\bibnamefont {Caballero}}, \bibinfo {author} {\bibfnamefont
  {M.}~\bibnamefont {Keith}}, \emph {et~al.},\ }\bibfield  {title} {\bibinfo
  {title} {The international pulsar timing array: first data release},\
  }\href@noop {} {\bibfield  {journal} {\bibinfo  {journal} {Monthly Notices of
  the Royal Astronomical Society}\ }\textbf {\bibinfo {volume} {458}},\
  \bibinfo {pages} {1267} (\bibinfo {year} {2016})}\BibitemShut {NoStop}%
\bibitem [{\citenamefont {An}(2019)}]{an2019science}%
  \BibitemOpen
  \bibfield  {author} {\bibinfo {author} {\bibfnamefont {T.}~\bibnamefont
  {An}},\ }\bibfield  {title} {\bibinfo {title} {Science opportunities and
  challenges associated with ska big data},\ }\href@noop {} {\bibfield
  {journal} {\bibinfo  {journal} {Science China Physics, Mechanics \&
  Astronomy}\ }\textbf {\bibinfo {volume} {62}},\ \bibinfo {pages} {1}
  (\bibinfo {year} {2019})}\BibitemShut {NoStop}%
\bibitem [{\citenamefont {Wang}\ \emph
  {et~al.}(2019{\natexlab{a}})\citenamefont {Wang}, \citenamefont {Pan},
  \citenamefont {Zheng}, \citenamefont {Qian},\ and\ \citenamefont
  {Li}}]{wang2019hybrid}%
  \BibitemOpen
  \bibfield  {author} {\bibinfo {author} {\bibfnamefont {Y.}~\bibnamefont
  {Wang}}, \bibinfo {author} {\bibfnamefont {Z.}~\bibnamefont {Pan}}, \bibinfo
  {author} {\bibfnamefont {J.}~\bibnamefont {Zheng}}, \bibinfo {author}
  {\bibfnamefont {L.}~\bibnamefont {Qian}},\ and\ \bibinfo {author}
  {\bibfnamefont {M.}~\bibnamefont {Li}},\ }\bibfield  {title} {\bibinfo
  {title} {A hybrid ensemble method for pulsar candidate classification},\
  }\href@noop {} {\bibfield  {journal} {\bibinfo  {journal} {Astrophysics and
  Space Science}\ }\textbf {\bibinfo {volume} {364}},\ \bibinfo {pages} {1}
  (\bibinfo {year} {2019}{\natexlab{a}})}\BibitemShut {NoStop}%
\bibitem [{\citenamefont {Tariq}\ \emph {et~al.}(2022)\citenamefont {Tariq},
  \citenamefont {Meng}, \citenamefont {Yao}, \citenamefont {Liu}, \citenamefont
  {Zhou}, \citenamefont {Ahmed},\ and\ \citenamefont
  {Spanakis-Misirlis}}]{tariq2022adaboost}%
  \BibitemOpen
  \bibfield  {author} {\bibinfo {author} {\bibfnamefont {I.}~\bibnamefont
  {Tariq}}, \bibinfo {author} {\bibfnamefont {Q.}~\bibnamefont {Meng}},
  \bibinfo {author} {\bibfnamefont {S.}~\bibnamefont {Yao}}, \bibinfo {author}
  {\bibfnamefont {W.}~\bibnamefont {Liu}}, \bibinfo {author} {\bibfnamefont
  {C.}~\bibnamefont {Zhou}}, \bibinfo {author} {\bibfnamefont {A.}~\bibnamefont
  {Ahmed}},\ and\ \bibinfo {author} {\bibfnamefont {A.}~\bibnamefont
  {Spanakis-Misirlis}},\ }\bibfield  {title} {\bibinfo {title} {Adaboost-dsnn:
  an adaptive boosting algorithm based on deep self normalized neural network
  for pulsar identification},\ }\href@noop {} {\bibfield  {journal} {\bibinfo
  {journal} {Monthly Notices of the Royal Astronomical Society}\ }\textbf
  {\bibinfo {volume} {511}},\ \bibinfo {pages} {683} (\bibinfo {year}
  {2022})}\BibitemShut {NoStop}%
\bibitem [{\citenamefont {Beniwal}\ \emph {et~al.}(2021)\citenamefont
  {Beniwal}, \citenamefont {Roy}, \citenamefont {Yadav},\ and\ \citenamefont
  {Chauhan}}]{beniwal2021detection}%
  \BibitemOpen
  \bibfield  {author} {\bibinfo {author} {\bibfnamefont {D.}~\bibnamefont
  {Beniwal}}, \bibinfo {author} {\bibfnamefont {A.}~\bibnamefont {Roy}},
  \bibinfo {author} {\bibfnamefont {H.}~\bibnamefont {Yadav}},\ and\ \bibinfo
  {author} {\bibfnamefont {A.}~\bibnamefont {Chauhan}},\ }\bibfield  {title}
  {\bibinfo {title} {Detection of pulsars by classical machine learning
  algorithms},\ }in\ \href@noop {} {\emph {\bibinfo {booktitle} {2021 2nd
  International Conference for Emerging Technology (INCET)}}}\ (\bibinfo
  {organization} {IEEE},\ \bibinfo {year} {2021})\ pp.\ \bibinfo {pages}
  {1--7}\BibitemShut {NoStop}%
\bibitem [{\citenamefont {Xiao}\ \emph {et~al.}(2020)\citenamefont {Xiao},
  \citenamefont {Li}, \citenamefont {Lin},\ and\ \citenamefont
  {Qiu}}]{xiao2020pulsar}%
  \BibitemOpen
  \bibfield  {author} {\bibinfo {author} {\bibfnamefont {J.}~\bibnamefont
  {Xiao}}, \bibinfo {author} {\bibfnamefont {X.}~\bibnamefont {Li}}, \bibinfo
  {author} {\bibfnamefont {H.}~\bibnamefont {Lin}},\ and\ \bibinfo {author}
  {\bibfnamefont {K.}~\bibnamefont {Qiu}},\ }\bibfield  {title} {\bibinfo
  {title} {Pulsar candidate selection using pseudo-nearest centroid neighbour
  classifier},\ }\href@noop {} {\bibfield  {journal} {\bibinfo  {journal}
  {Monthly Notices of the Royal Astronomical Society}\ }\textbf {\bibinfo
  {volume} {492}},\ \bibinfo {pages} {2119} (\bibinfo {year}
  {2020})}\BibitemShut {NoStop}%
\bibitem [{\citenamefont {Lee}\ \emph {et~al.}(2012)\citenamefont {Lee},
  \citenamefont {Guillemot}, \citenamefont {Yue}, \citenamefont {Kramer},\ and\
  \citenamefont {Champion}}]{lee2012application}%
  \BibitemOpen
  \bibfield  {author} {\bibinfo {author} {\bibfnamefont {K.}~\bibnamefont
  {Lee}}, \bibinfo {author} {\bibfnamefont {L.}~\bibnamefont {Guillemot}},
  \bibinfo {author} {\bibfnamefont {Y.}~\bibnamefont {Yue}}, \bibinfo {author}
  {\bibfnamefont {M.}~\bibnamefont {Kramer}},\ and\ \bibinfo {author}
  {\bibfnamefont {D.}~\bibnamefont {Champion}},\ }\bibfield  {title} {\bibinfo
  {title} {Application of the gaussian mixture model in pulsar astronomy-pulsar
  classification and candidates ranking for the fermi 2fgl catalogue},\
  }\href@noop {} {\bibfield  {journal} {\bibinfo  {journal} {Monthly Notices of
  the Royal Astronomical Society}\ }\textbf {\bibinfo {volume} {424}},\
  \bibinfo {pages} {2832} (\bibinfo {year} {2012})}\BibitemShut {NoStop}%
\bibitem [{\citenamefont {Wang}\ \emph
  {et~al.}(2019{\natexlab{b}})\citenamefont {Wang}, \citenamefont {Li},
  \citenamefont {Pan},\ and\ \citenamefont {Zheng}}]{wang2019pulsar}%
  \BibitemOpen
  \bibfield  {author} {\bibinfo {author} {\bibfnamefont {Y.-C.}\ \bibnamefont
  {Wang}}, \bibinfo {author} {\bibfnamefont {M.-T.}\ \bibnamefont {Li}},
  \bibinfo {author} {\bibfnamefont {Z.-C.}\ \bibnamefont {Pan}},\ and\ \bibinfo
  {author} {\bibfnamefont {J.-H.}\ \bibnamefont {Zheng}},\ }\bibfield  {title}
  {\bibinfo {title} {Pulsar candidate classification with deep convolutional
  neural networks},\ }\href@noop {} {\bibfield  {journal} {\bibinfo  {journal}
  {Research in Astronomy and Astrophysics}\ }\textbf {\bibinfo {volume} {19}},\
  \bibinfo {pages} {133} (\bibinfo {year} {2019}{\natexlab{b}})}\BibitemShut
  {NoStop}%
\bibitem [{\citenamefont {Lyon}\ \emph {et~al.}(2016)\citenamefont {Lyon},
  \citenamefont {Stappers}, \citenamefont {Cooper}, \citenamefont {Brooke},\
  and\ \citenamefont {Knowles}}]{lyon2016fifty}%
  \BibitemOpen
  \bibfield  {author} {\bibinfo {author} {\bibfnamefont {R.~J.}\ \bibnamefont
  {Lyon}}, \bibinfo {author} {\bibfnamefont {B.}~\bibnamefont {Stappers}},
  \bibinfo {author} {\bibfnamefont {S.}~\bibnamefont {Cooper}}, \bibinfo
  {author} {\bibfnamefont {J.~M.}\ \bibnamefont {Brooke}},\ and\ \bibinfo
  {author} {\bibfnamefont {J.~D.}\ \bibnamefont {Knowles}},\ }\bibfield
  {title} {\bibinfo {title} {Fifty years of pulsar candidate selection: from
  simple filters to a new principled real-time classification approach},\
  }\href@noop {} {\bibfield  {journal} {\bibinfo  {journal} {Monthly Notices of
  the Royal Astronomical Society}\ }\textbf {\bibinfo {volume} {459}},\
  \bibinfo {pages} {1104} (\bibinfo {year} {2016})}\BibitemShut {NoStop}%
\bibitem [{\citenamefont {Zhou}\ \emph {et~al.}(2017)\citenamefont {Zhou},
  \citenamefont {Pan}, \citenamefont {Wang},\ and\ \citenamefont
  {Vasilakos}}]{zhou2017machine}%
  \BibitemOpen
  \bibfield  {author} {\bibinfo {author} {\bibfnamefont {L.}~\bibnamefont
  {Zhou}}, \bibinfo {author} {\bibfnamefont {S.}~\bibnamefont {Pan}}, \bibinfo
  {author} {\bibfnamefont {J.}~\bibnamefont {Wang}},\ and\ \bibinfo {author}
  {\bibfnamefont {A.~V.}\ \bibnamefont {Vasilakos}},\ }\bibfield  {title}
  {\bibinfo {title} {Machine learning on big data: Opportunities and
  challenges},\ }\href@noop {} {\bibfield  {journal} {\bibinfo  {journal}
  {Neurocomputing}\ }\textbf {\bibinfo {volume} {237}},\ \bibinfo {pages} {350}
  (\bibinfo {year} {2017})}\BibitemShut {NoStop}%
\bibitem [{\citenamefont {Sarma}\ \emph {et~al.}(2019)\citenamefont {Sarma},
  \citenamefont {Chatterjee}, \citenamefont {Gili},\ and\ \citenamefont
  {Yu}}]{sarma2019quantum}%
  \BibitemOpen
  \bibfield  {author} {\bibinfo {author} {\bibfnamefont {A.}~\bibnamefont
  {Sarma}}, \bibinfo {author} {\bibfnamefont {R.}~\bibnamefont {Chatterjee}},
  \bibinfo {author} {\bibfnamefont {K.}~\bibnamefont {Gili}},\ and\ \bibinfo
  {author} {\bibfnamefont {T.}~\bibnamefont {Yu}},\ }\bibfield  {title}
  {\bibinfo {title} {Quantum unsupervised and supervised learning on
  superconducting processors},\ }\href@noop {} {\bibfield  {journal} {\bibinfo
  {journal} {arXiv preprint arXiv:1909.04226}\ } (\bibinfo {year}
  {2019})}\BibitemShut {NoStop}%
\bibitem [{\citenamefont {Kordzanganeh}\ \emph {et~al.}(2021)\citenamefont
  {Kordzanganeh}, \citenamefont {Utting},\ and\ \citenamefont
  {Scaife}}]{kordzanganeh2021quantum}%
  \BibitemOpen
  \bibfield  {author} {\bibinfo {author} {\bibfnamefont {M.}~\bibnamefont
  {Kordzanganeh}}, \bibinfo {author} {\bibfnamefont {A.}~\bibnamefont
  {Utting}},\ and\ \bibinfo {author} {\bibfnamefont {A.}~\bibnamefont
  {Scaife}},\ }\bibfield  {title} {\bibinfo {title} {Quantum machine learning
  for radio astronomy},\ }\href@noop {} {\bibfield  {journal} {\bibinfo
  {journal} {arXiv preprint arXiv:2112.02655}\ } (\bibinfo {year}
  {2021})}\BibitemShut {NoStop}%
\bibitem [{\citenamefont {Lyon}(2017)}]{misc_htru2_372}%
  \BibitemOpen
  \bibfield  {author} {\bibinfo {author} {\bibfnamefont {R.}~\bibnamefont
  {Lyon}},\ }\href@noop {} {\bibinfo {title} {{HTRU2}}},\ \bibinfo
  {howpublished} {UCI Machine Learning Repository} (\bibinfo {year} {2017}),\
  \bibinfo {note} {{DOI}: https://doi.org/10.24432/C5DK6R}\BibitemShut
  {NoStop}%
\bibitem [{\citenamefont {Mohri}\ \emph {et~al.}(2018)\citenamefont {Mohri},
  \citenamefont {Rostamizadeh},\ and\ \citenamefont
  {Talwalkar}}]{mohri2018foundations}%
  \BibitemOpen
  \bibfield  {author} {\bibinfo {author} {\bibfnamefont {M.}~\bibnamefont
  {Mohri}}, \bibinfo {author} {\bibfnamefont {A.}~\bibnamefont
  {Rostamizadeh}},\ and\ \bibinfo {author} {\bibfnamefont {A.}~\bibnamefont
  {Talwalkar}},\ }\href@noop {} {\emph {\bibinfo {title} {Foundations of
  machine learning}}}\ (\bibinfo  {publisher} {MIT press},\ \bibinfo {year}
  {2018})\BibitemShut {NoStop}%
\bibitem [{\citenamefont {Alpaydin}(2020)}]{alpaydin2020introduction}%
  \BibitemOpen
  \bibfield  {author} {\bibinfo {author} {\bibfnamefont {E.}~\bibnamefont
  {Alpaydin}},\ }\href@noop {} {\emph {\bibinfo {title} {Introduction to
  machine learning}}}\ (\bibinfo  {publisher} {MIT press},\ \bibinfo {year}
  {2020})\BibitemShut {NoStop}%
\bibitem [{\citenamefont {Schuld}(2021{\natexlab{a}})}]{schuld2021supervised}%
  \BibitemOpen
  \bibfield  {author} {\bibinfo {author} {\bibfnamefont {M.}~\bibnamefont
  {Schuld}},\ }\bibfield  {title} {\bibinfo {title} {Supervised quantum machine
  learning models are kernel methods},\ }\href@noop {} {\bibfield  {journal}
  {\bibinfo  {journal} {arXiv preprint arXiv:2101.11020}\ } (\bibinfo {year}
  {2021}{\natexlab{a}})}\BibitemShut {NoStop}%
\bibitem [{\citenamefont {M{\"u}ller}\ and\ \citenamefont
  {Guido}(2016)}]{muller2016introduction}%
  \BibitemOpen
  \bibfield  {author} {\bibinfo {author} {\bibfnamefont {A.~C.}\ \bibnamefont
  {M{\"u}ller}}\ and\ \bibinfo {author} {\bibfnamefont {S.}~\bibnamefont
  {Guido}},\ }\href@noop {} {\emph {\bibinfo {title} {Introduction to machine
  learning with Python: a guide for data scientists}}}\ (\bibinfo  {publisher}
  {``O'Reilly Media, Inc."},\ \bibinfo {year} {2016})\BibitemShut {NoStop}%
\bibitem [{\citenamefont {Powers}(2020)}]{powers2020evaluation}%
  \BibitemOpen
  \bibfield  {author} {\bibinfo {author} {\bibfnamefont {D.~M.}\ \bibnamefont
  {Powers}},\ }\bibfield  {title} {\bibinfo {title} {Evaluation: from
  precision, recall and f-measure to roc, informedness, markedness and
  correlation},\ }\href@noop {} {\bibfield  {journal} {\bibinfo  {journal}
  {arXiv preprint arXiv:2010.16061}\ } (\bibinfo {year} {2020})}\BibitemShut
  {NoStop}%
\bibitem [{\citenamefont {Nielsen}\ and\ \citenamefont
  {Chuang}(2001)}]{nielsen2001quantum}%
  \BibitemOpen
  \bibfield  {author} {\bibinfo {author} {\bibfnamefont {M.~A.}\ \bibnamefont
  {Nielsen}}\ and\ \bibinfo {author} {\bibfnamefont {I.~L.}\ \bibnamefont
  {Chuang}},\ }\bibfield  {title} {\bibinfo {title} {Quantum computation and
  quantum information},\ }\href@noop {} {\bibfield  {journal} {\bibinfo
  {journal} {Phys. Today}\ }\textbf {\bibinfo {volume} {54}},\ \bibinfo {pages}
  {60} (\bibinfo {year} {2001})}\BibitemShut {NoStop}%
\bibitem [{\citenamefont {Schuld}\ and\ \citenamefont
  {Petruccione}(2021)}]{schuld2021machine}%
  \BibitemOpen
  \bibfield  {author} {\bibinfo {author} {\bibfnamefont {M.}~\bibnamefont
  {Schuld}}\ and\ \bibinfo {author} {\bibfnamefont {F.}~\bibnamefont
  {Petruccione}},\ }\href@noop {} {\emph {\bibinfo {title} {Machine learning
  with quantum computers}}}\ (\bibinfo  {publisher} {Springer},\ \bibinfo
  {year} {2021})\BibitemShut {NoStop}%
\bibitem [{\citenamefont {Tharwat}(2020)}]{tharwat2020classification}%
  \BibitemOpen
  \bibfield  {author} {\bibinfo {author} {\bibfnamefont {A.}~\bibnamefont
  {Tharwat}},\ }\bibfield  {title} {\bibinfo {title} {Classification assessment
  methods},\ }\href@noop {} {\bibfield  {journal} {\bibinfo  {journal} {Applied
  computing and informatics}\ }\textbf {\bibinfo {volume} {17}},\ \bibinfo
  {pages} {168} (\bibinfo {year} {2020})}\BibitemShut {NoStop}%
\bibitem [{\citenamefont {Noble}(2006)}]{noble2006support}%
  \BibitemOpen
  \bibfield  {author} {\bibinfo {author} {\bibfnamefont {W.~S.}\ \bibnamefont
  {Noble}},\ }\bibfield  {title} {\bibinfo {title} {What is a support vector
  machine?},\ }\href@noop {} {\bibfield  {journal} {\bibinfo  {journal} {Nature
  biotechnology}\ }\textbf {\bibinfo {volume} {24}},\ \bibinfo {pages} {1565}
  (\bibinfo {year} {2006})}\BibitemShut {NoStop}%
\bibitem [{\citenamefont {Boser}\ \emph {et~al.}(1992)\citenamefont {Boser},
  \citenamefont {Guyon},\ and\ \citenamefont {Vapnik}}]{boser1992training}%
  \BibitemOpen
  \bibfield  {author} {\bibinfo {author} {\bibfnamefont {B.~E.}\ \bibnamefont
  {Boser}}, \bibinfo {author} {\bibfnamefont {I.~M.}\ \bibnamefont {Guyon}},\
  and\ \bibinfo {author} {\bibfnamefont {V.~N.}\ \bibnamefont {Vapnik}},\
  }\bibfield  {title} {\bibinfo {title} {A training algorithm for optimal
  margin classifiers},\ }in\ \href@noop {} {\emph {\bibinfo {booktitle}
  {Proceedings of the fifth annual workshop on Computational learning
  theory}}}\ (\bibinfo {year} {1992})\ pp.\ \bibinfo {pages}
  {144--152}\BibitemShut {NoStop}%
\bibitem [{\citenamefont {Sch{\"o}lkopf}\ \emph {et~al.}(1999)\citenamefont
  {Sch{\"o}lkopf}, \citenamefont {Burges},\ and\ \citenamefont
  {Smola}}]{scholkopf1999advances}%
  \BibitemOpen
  \bibfield  {author} {\bibinfo {author} {\bibfnamefont {B.}~\bibnamefont
  {Sch{\"o}lkopf}}, \bibinfo {author} {\bibfnamefont {C.~J.}\ \bibnamefont
  {Burges}},\ and\ \bibinfo {author} {\bibfnamefont {A.~J.}\ \bibnamefont
  {Smola}},\ }\href@noop {} {\emph {\bibinfo {title} {Advances in kernel
  methods: support vector learning}}}\ (\bibinfo  {publisher} {MIT press},\
  \bibinfo {year} {1999})\BibitemShut {NoStop}%
\bibitem [{\citenamefont {Havl{\'\i}{\v{c}}ek}\ \emph
  {et~al.}(2019)\citenamefont {Havl{\'\i}{\v{c}}ek}, \citenamefont
  {C{\'o}rcoles}, \citenamefont {Temme}, \citenamefont {Harrow}, \citenamefont
  {Kandala}, \citenamefont {Chow},\ and\ \citenamefont
  {Gambetta}}]{havlivcek2019supervised}%
  \BibitemOpen
  \bibfield  {author} {\bibinfo {author} {\bibfnamefont {V.}~\bibnamefont
  {Havl{\'\i}{\v{c}}ek}}, \bibinfo {author} {\bibfnamefont {A.~D.}\
  \bibnamefont {C{\'o}rcoles}}, \bibinfo {author} {\bibfnamefont
  {K.}~\bibnamefont {Temme}}, \bibinfo {author} {\bibfnamefont {A.~W.}\
  \bibnamefont {Harrow}}, \bibinfo {author} {\bibfnamefont {A.}~\bibnamefont
  {Kandala}}, \bibinfo {author} {\bibfnamefont {J.~M.}\ \bibnamefont {Chow}},\
  and\ \bibinfo {author} {\bibfnamefont {J.~M.}\ \bibnamefont {Gambetta}},\
  }\bibfield  {title} {\bibinfo {title} {Supervised learning with
  quantum-enhanced feature spaces},\ }\href@noop {} {\bibfield  {journal}
  {\bibinfo  {journal} {Nature}\ }\textbf {\bibinfo {volume} {567}},\ \bibinfo
  {pages} {209} (\bibinfo {year} {2019})}\BibitemShut {NoStop}%
\bibitem [{\citenamefont {Schuld}\ and\ \citenamefont
  {Killoran}(2019)}]{schuld2019quantum}%
  \BibitemOpen
  \bibfield  {author} {\bibinfo {author} {\bibfnamefont {M.}~\bibnamefont
  {Schuld}}\ and\ \bibinfo {author} {\bibfnamefont {N.}~\bibnamefont
  {Killoran}},\ }\bibfield  {title} {\bibinfo {title} {Quantum machine learning
  in feature hilbert spaces},\ }\href@noop {} {\bibfield  {journal} {\bibinfo
  {journal} {Physical review letters}\ }\textbf {\bibinfo {volume} {122}},\
  \bibinfo {pages} {040504} (\bibinfo {year} {2019})}\BibitemShut {NoStop}%
\bibitem [{\citenamefont {Hubregtsen}\ \emph {et~al.}(2022)\citenamefont
  {Hubregtsen}, \citenamefont {Wierichs}, \citenamefont {Gil-Fuster},
  \citenamefont {Derks}, \citenamefont {Faehrmann},\ and\ \citenamefont
  {Meyer}}]{hubregtsen2022training}%
  \BibitemOpen
  \bibfield  {author} {\bibinfo {author} {\bibfnamefont {T.}~\bibnamefont
  {Hubregtsen}}, \bibinfo {author} {\bibfnamefont {D.}~\bibnamefont
  {Wierichs}}, \bibinfo {author} {\bibfnamefont {E.}~\bibnamefont
  {Gil-Fuster}}, \bibinfo {author} {\bibfnamefont {P.-J.~H.}\ \bibnamefont
  {Derks}}, \bibinfo {author} {\bibfnamefont {P.~K.}\ \bibnamefont
  {Faehrmann}},\ and\ \bibinfo {author} {\bibfnamefont {J.~J.}\ \bibnamefont
  {Meyer}},\ }\bibfield  {title} {\bibinfo {title} {Training quantum embedding
  kernels on near-term quantum computers},\ }\href@noop {} {\bibfield
  {journal} {\bibinfo  {journal} {Physical Review A}\ }\textbf {\bibinfo
  {volume} {106}},\ \bibinfo {pages} {042431} (\bibinfo {year}
  {2022})}\BibitemShut {NoStop}%
\bibitem [{\citenamefont {Krizhevsky}\ \emph {et~al.}(2012)\citenamefont
  {Krizhevsky}, \citenamefont {Sutskever},\ and\ \citenamefont
  {Hinton}}]{krizhevsky2012imagenet}%
  \BibitemOpen
  \bibfield  {author} {\bibinfo {author} {\bibfnamefont {A.}~\bibnamefont
  {Krizhevsky}}, \bibinfo {author} {\bibfnamefont {I.}~\bibnamefont
  {Sutskever}},\ and\ \bibinfo {author} {\bibfnamefont {G.~E.}\ \bibnamefont
  {Hinton}},\ }\bibfield  {title} {\bibinfo {title} {Imagenet classification
  with deep convolutional neural networks},\ }\href@noop {} {\bibfield
  {journal} {\bibinfo  {journal} {Advances in neural information processing
  systems}\ }\textbf {\bibinfo {volume} {25}} (\bibinfo {year}
  {2012})}\BibitemShut {NoStop}%
\bibitem [{\citenamefont {Gu}\ \emph {et~al.}(2018)\citenamefont {Gu},
  \citenamefont {Wang}, \citenamefont {Kuen}, \citenamefont {Ma}, \citenamefont
  {Shahroudy}, \citenamefont {Shuai}, \citenamefont {Liu}, \citenamefont
  {Wang}, \citenamefont {Wang}, \citenamefont {Cai} \emph
  {et~al.}}]{gu2018recent}%
  \BibitemOpen
  \bibfield  {author} {\bibinfo {author} {\bibfnamefont {J.}~\bibnamefont
  {Gu}}, \bibinfo {author} {\bibfnamefont {Z.}~\bibnamefont {Wang}}, \bibinfo
  {author} {\bibfnamefont {J.}~\bibnamefont {Kuen}}, \bibinfo {author}
  {\bibfnamefont {L.}~\bibnamefont {Ma}}, \bibinfo {author} {\bibfnamefont
  {A.}~\bibnamefont {Shahroudy}}, \bibinfo {author} {\bibfnamefont
  {B.}~\bibnamefont {Shuai}}, \bibinfo {author} {\bibfnamefont
  {T.}~\bibnamefont {Liu}}, \bibinfo {author} {\bibfnamefont {X.}~\bibnamefont
  {Wang}}, \bibinfo {author} {\bibfnamefont {G.}~\bibnamefont {Wang}}, \bibinfo
  {author} {\bibfnamefont {J.}~\bibnamefont {Cai}}, \emph {et~al.},\ }\bibfield
   {title} {\bibinfo {title} {Recent advances in convolutional neural
  networks},\ }\href@noop {} {\bibfield  {journal} {\bibinfo  {journal}
  {Pattern recognition}\ }\textbf {\bibinfo {volume} {77}},\ \bibinfo {pages}
  {354} (\bibinfo {year} {2018})}\BibitemShut {NoStop}%
\bibitem [{\citenamefont {Hur}\ \emph {et~al.}(2022)\citenamefont {Hur},
  \citenamefont {Kim},\ and\ \citenamefont {Park}}]{hur2022quantum}%
  \BibitemOpen
  \bibfield  {author} {\bibinfo {author} {\bibfnamefont {T.}~\bibnamefont
  {Hur}}, \bibinfo {author} {\bibfnamefont {L.}~\bibnamefont {Kim}},\ and\
  \bibinfo {author} {\bibfnamefont {D.~K.}\ \bibnamefont {Park}},\ }\bibfield
  {title} {\bibinfo {title} {Quantum convolutional neural network for classical
  data classification},\ }\href@noop {} {\bibfield  {journal} {\bibinfo
  {journal} {Quantum Machine Intelligence}\ }\textbf {\bibinfo {volume} {4}},\
  \bibinfo {pages} {3} (\bibinfo {year} {2022})}\BibitemShut {NoStop}%
\bibitem [{\citenamefont {Henderson}\ \emph {et~al.}(2020)\citenamefont
  {Henderson}, \citenamefont {Shakya}, \citenamefont {Pradhan},\ and\
  \citenamefont {Cook}}]{henderson2020quanvolutional}%
  \BibitemOpen
  \bibfield  {author} {\bibinfo {author} {\bibfnamefont {M.}~\bibnamefont
  {Henderson}}, \bibinfo {author} {\bibfnamefont {S.}~\bibnamefont {Shakya}},
  \bibinfo {author} {\bibfnamefont {S.}~\bibnamefont {Pradhan}},\ and\ \bibinfo
  {author} {\bibfnamefont {T.}~\bibnamefont {Cook}},\ }\bibfield  {title}
  {\bibinfo {title} {Quanvolutional neural networks: powering image recognition
  with quantum circuits},\ }\href@noop {} {\bibfield  {journal} {\bibinfo
  {journal} {Quantum Machine Intelligence}\ }\textbf {\bibinfo {volume} {2}},\
  \bibinfo {pages} {2} (\bibinfo {year} {2020})}\BibitemShut {NoStop}%
\bibitem [{\citenamefont {Wei}\ \emph {et~al.}(2022)\citenamefont {Wei},
  \citenamefont {Chen}, \citenamefont {Zhou},\ and\ \citenamefont
  {Long}}]{wei2022quantum}%
  \BibitemOpen
  \bibfield  {author} {\bibinfo {author} {\bibfnamefont {S.}~\bibnamefont
  {Wei}}, \bibinfo {author} {\bibfnamefont {Y.}~\bibnamefont {Chen}}, \bibinfo
  {author} {\bibfnamefont {Z.}~\bibnamefont {Zhou}},\ and\ \bibinfo {author}
  {\bibfnamefont {G.}~\bibnamefont {Long}},\ }\bibfield  {title} {\bibinfo
  {title} {A quantum convolutional neural network on nisq devices},\
  }\href@noop {} {\bibfield  {journal} {\bibinfo  {journal} {AAPPS Bulletin}\
  }\textbf {\bibinfo {volume} {32}},\ \bibinfo {pages} {1} (\bibinfo {year}
  {2022})}\BibitemShut {NoStop}%
\bibitem [{\citenamefont {Cong}\ \emph {et~al.}(2019)\citenamefont {Cong},
  \citenamefont {Choi},\ and\ \citenamefont {Lukin}}]{cong2019quantum}%
  \BibitemOpen
  \bibfield  {author} {\bibinfo {author} {\bibfnamefont {I.}~\bibnamefont
  {Cong}}, \bibinfo {author} {\bibfnamefont {S.}~\bibnamefont {Choi}},\ and\
  \bibinfo {author} {\bibfnamefont {M.~D.}\ \bibnamefont {Lukin}},\ }\bibfield
  {title} {\bibinfo {title} {Quantum convolutional neural networks},\
  }\href@noop {} {\bibfield  {journal} {\bibinfo  {journal} {Nature Physics}\
  }\textbf {\bibinfo {volume} {15}},\ \bibinfo {pages} {1273} (\bibinfo {year}
  {2019})}\BibitemShut {NoStop}%
\bibitem [{\citenamefont {Lourens}\ \emph {et~al.}(2023)\citenamefont
  {Lourens}, \citenamefont {Sinayskiy}, \citenamefont {Park}, \citenamefont
  {Blank},\ and\ \citenamefont {Petruccione}}]{lourens2023hierarchical}%
  \BibitemOpen
  \bibfield  {author} {\bibinfo {author} {\bibfnamefont {M.}~\bibnamefont
  {Lourens}}, \bibinfo {author} {\bibfnamefont {I.}~\bibnamefont {Sinayskiy}},
  \bibinfo {author} {\bibfnamefont {D.~K.}\ \bibnamefont {Park}}, \bibinfo
  {author} {\bibfnamefont {C.}~\bibnamefont {Blank}},\ and\ \bibinfo {author}
  {\bibfnamefont {F.}~\bibnamefont {Petruccione}},\ }\bibfield  {title}
  {\bibinfo {title} {Hierarchical quantum circuit representations for neural
  architecture search},\ }\href@noop {} {\bibfield  {journal} {\bibinfo
  {journal} {npj Quantum Information}\ }\textbf {\bibinfo {volume} {9}},\
  \bibinfo {pages} {79} (\bibinfo {year} {2023})}\BibitemShut {NoStop}%
\bibitem [{\citenamefont {Schuld}\ \emph {et~al.}(2021)\citenamefont {Schuld},
  \citenamefont {Sweke},\ and\ \citenamefont {Meyer}}]{schuld2021effect}%
  \BibitemOpen
  \bibfield  {author} {\bibinfo {author} {\bibfnamefont {M.}~\bibnamefont
  {Schuld}}, \bibinfo {author} {\bibfnamefont {R.}~\bibnamefont {Sweke}},\ and\
  \bibinfo {author} {\bibfnamefont {J.~J.}\ \bibnamefont {Meyer}},\ }\bibfield
  {title} {\bibinfo {title} {Effect of data encoding on the expressive power of
  variational quantum-machine-learning models},\ }\href@noop {} {\bibfield
  {journal} {\bibinfo  {journal} {Physical Review A}\ }\textbf {\bibinfo
  {volume} {103}},\ \bibinfo {pages} {032430} (\bibinfo {year}
  {2021})}\BibitemShut {NoStop}%
\bibitem [{\citenamefont {Lyon}(2016)}]{lyon2016pulsars}%
  \BibitemOpen
  \bibfield  {author} {\bibinfo {author} {\bibfnamefont {R.~J.}\ \bibnamefont
  {Lyon}},\ }\href@noop {} {\emph {\bibinfo {title} {Why are pulsars hard to
  find?}}}\ (\bibinfo  {publisher} {The University of Manchester (United
  Kingdom)},\ \bibinfo {year} {2016})\BibitemShut {NoStop}%
\bibitem [{\citenamefont {Schuld}(2021{\natexlab{b}})}]{Schuld_2021_demo}%
  \BibitemOpen
  \bibfield  {author} {\bibinfo {author} {\bibfnamefont {M.}~\bibnamefont
  {Schuld}},\ }\href
  {https://pennylane.ai/qml/demos/tutorial_kernel_based_training} {\bibinfo
  {title} {Kernel-based training of quantum models with scikit-learn}}
  (\bibinfo {year} {2021}{\natexlab{b}})\BibitemShut {NoStop}%
\bibitem [{qis(2023)}]{qiskit_runtime}%
  \BibitemOpen
  \href@noop {} {\bibinfo {title} {Qiskit runtime}},\ \bibinfo {howpublished}
  {\url{https://qiskit.org/ecosystem/ibm-runtime/index.html}} (\bibinfo {year}
  {2023}),\ \bibinfo {note} {version 0.12.0, Accessed on
  09/08/2023}\BibitemShut {NoStop}%
\bibitem [{\citenamefont {{Qiskit contributors}}(2023)}]{Qiskit}%
  \BibitemOpen
  \bibfield  {author} {\bibinfo {author} {\bibnamefont {{Qiskit
  contributors}}},\ }\href {https://doi.org/10.5281/zenodo.2573505} {\bibinfo
  {title} {Qiskit: An open-source framework for quantum computing}} (\bibinfo
  {year} {2023})\BibitemShut {NoStop}%
\bibitem [{\citenamefont {Bergholm}\ \emph {et~al.}(2018)\citenamefont
  {Bergholm}, \citenamefont {Izaac}, \citenamefont {Schuld}, \citenamefont
  {Gogolin}, \citenamefont {Blank}, \citenamefont {McKiernan},\ and\
  \citenamefont {Killoran}}]{bergholm2018pennylane}%
  \BibitemOpen
  \bibfield  {author} {\bibinfo {author} {\bibfnamefont {V.}~\bibnamefont
  {Bergholm}}, \bibinfo {author} {\bibfnamefont {J.}~\bibnamefont {Izaac}},
  \bibinfo {author} {\bibfnamefont {M.}~\bibnamefont {Schuld}}, \bibinfo
  {author} {\bibfnamefont {C.}~\bibnamefont {Gogolin}}, \bibinfo {author}
  {\bibfnamefont {C.}~\bibnamefont {Blank}}, \bibinfo {author} {\bibfnamefont
  {K.}~\bibnamefont {McKiernan}},\ and\ \bibinfo {author} {\bibfnamefont
  {N.}~\bibnamefont {Killoran}},\ }\bibfield  {title} {\bibinfo {title}
  {Pennylane},\ }\href@noop {} {\bibfield  {journal} {\bibinfo  {journal}
  {arXiv}\ } (\bibinfo {year} {2018})},\ \Eprint
  {https://arxiv.org/abs/1811.04968} {arXiv:1811.04968 [quant-ph]} \BibitemShut
  {NoStop}%
\bibitem [{\citenamefont {{IBM Quantum
  Experience}}(2023)}]{IBMQuantumPlatform}%
  \BibitemOpen
  \bibfield  {author} {\bibinfo {author} {\bibnamefont {{IBM Quantum
  Experience}}},\ }\href@noop {} {\bibinfo {title} {{IBM Quantum Platform}}},\
  \bibinfo {howpublished} {\url{https://quantum-computing.ibm.com/}} (\bibinfo
  {year} {2023}),\ \bibinfo {note} {accessed on 2023/09/07}\BibitemShut
  {NoStop}%
\end{thebibliography}%

\end{document}